%% file: mapping_main.tex
\pdfoutput=1
\documentclass[10pt]{article}

\usepackage{cite}
\usepackage{amsmath,amssymb,amsfonts}
\usepackage{graphicx}
\usepackage{textcomp}
\usepackage{xcolor}
\def\BibTeX{{\rm B\kern-.05em{\sc i\kern-.025em b}\kern-.08em
    T\kern-.1667em\lower.7ex\hbox{E}\kern-.125emX}}

\usepackage[most]{tcolorbox}
\usepackage{hyperref}

\colorlet{punct}{red!60!black}
\definecolor{background}{HTML}{EEEEEE}
\definecolor{delim}{RGB}{20,105,176}
\colorlet{numb}{magenta!60!black}

\lstdefinelanguage{json}{
    basicstyle=\normalfont\ttfamily,
    numbers=left,
    numberstyle=\scriptsize,
    stepnumber=1,
    numbersep=8pt,
    showstringspaces=false,
    breaklines=true,
    frame=lines,
    backgroundcolor=\color{background},
    literate=
     *{0}{{{\color{numb}0}}}{1}
      {1}{{{\color{numb}1}}}{1}
      {2}{{{\color{numb}2}}}{1}
      {3}{{{\color{numb}3}}}{1}
      {4}{{{\color{numb}4}}}{1}
      {5}{{{\color{numb}5}}}{1}
      {6}{{{\color{numb}6}}}{1}
      {7}{{{\color{numb}7}}}{1}
      {8}{{{\color{numb}8}}}{1}
      {9}{{{\color{numb}9}}}{1}
      {:}{{{\color{punct}{:}}}}{1}
      {,}{{{\color{punct}{,}}}}{1}
      {\{}{{{\color{delim}{\{}}}}{1}
      {\}}{{{\color{delim}{\}}}}}{1}
      {[}{{{\color{delim}{[}}}}{1}
      {]}{{{\color{delim}{]}}}}{1},
}
\providecommand{\keywords}[1]
{
	\small	
	\textbf{\textit{Keywords---}} #1
}
\usepackage{listings}

\hypersetup{
	colorlinks=true, 
	linktoc=all,     
	linkcolor={black},
	urlcolor={blue},
	anchorcolor=black, 
	citecolor=black,
}

\usepackage{etoolbox}

\usepackage{diagbox} 
\usepackage[normalem]{ulem}
\usepackage{cooltooltips} 
\usepackage{subfigure}
\usepackage[disable]{todonotes} 
\usepackage{xspace} 
\usepackage{framed} 
\usepackage{mdframed}
\usepackage{booktabs}
\usepackage{multirow}
\usepackage[alsoload=binary,binary-units]{siunitx}
\usepackage{authblk}

\definecolor{jonas}{RGB}{128, 1, 0}

\newcommand{\FC}[1]{{\color{black}#1}}

\newcommand{\LP}[1]{{\color{black}#1}}

\newcommand{\JK}[1]{{\color{black}#1}}

\newcommand{\cmcount}{\textit{commMatrix count}\xspace}
\newcommand{\cmsize}{\textit{commMatrix size}\xspace}
\newcommand{\maplib}{\textit{MapLib}\xspace}
\newcommand{\ncdr}{NCD$_r$\xspace}

\definecolor{sweepgreen}{rgb}{0.22, 0.62, 0.2}
\definecolor{flowred}{HTML}{ea4141}
\definecolor{flowblue}{HTML}{366abf}
\definecolor{floworange}{HTML}{ce8437}
\definecolor{flowgreen}{HTML}{53ab2e}
\definecolor{olive}{HTML}{cddf31}

\begin{document}
	\pagestyle{plain}
	\sloppy
%
%
%
%
%
\title{Mapping Matters: Application Process Mapping on 3-D Processor Topologies
}

	\author[1]{Jonas H. M\"uller Kornd\"orfer}
	\author[2]{Mario Bielert}
	\author[3]{La\'ercio L. Pilla}
	\author[1]{Florina M. Ciorba}
	\affil[1]{
		Department of Mathematics and Computer Science\\

		University of Basel, Basel, Switzerland\\

		Email: jonas.korndorfer@unibas.ch and florina.ciorba@unibas.ch
	}
	\affil[2]{
		Center for Information Services and High Performance Computing (ZIH)\\

		Technische Universit\"at Dresden, Dresden, Germany\\

		Email: mario.bielert@tu-dresden.de
	}
	\affil[3]{
		Laboratoire de Recherche en Informatique, CNRS \&

		Univ. Paris-Saclay, Orsay, France\\

		Email: pilla@lri.fr
	}

\maketitle


\begin{abstract}
Applications' performance is influenced by the mapping of processes to computing nodes, the frequency and volume of exchanges among processing elements, the network capacity, and the routing protocol.
A poor mapping of application processes degrades performance and wastes system resources.
Process mapping is frequently ignored as an explicit optimization step since the system typically offers a default mapping, users may lack awareness of their applications' communication behavior, making the opportunities for improving performance through careful mapping are often unclear.
This work studies the impact of application process mapping on several processor topologies.
We propose a workflow that renders mapping as an explicit optimization step for parallel applications.
We apply the workflow to a set of four applications, twelve mapping algorithms, and three direct network topologies.
We assess the mappings' quality in terms of volume, frequency, and distance of exchanges using metrics such as dilation (measured in hop$\cdot$Byte).
With a parallel trace-based simulator, we predict the applications' execution on the three topologies using the twelve mappings.
We evaluate the impact of process mapping on the applications' simulated performance in terms of execution and communication times and identify the mappings that achieve the highest performance in both cases.
To ensure the correctness of the simulations, we compare the pre- and post-simulation results.
This work emphasizes the importance of process mapping as an explicit optimization step and, thereby, offers a solution for parallel applications to exploit the full potential of the allocated resources on a given system.
\end{abstract}


%
%

\input{01-introduction}
\input{02-relatedWork}
\input{03-proposedApproach}

\input{04-applications}
\input{05-topologies}
\input{06-core}
\input{07-experiments}
\input{08-conclusion}
\input{09-acknowledgment}

\bibliographystyle{acm}
\bibliography{mappingPaper-bibliography}

\newpage

\input{10-appendix}

\end{document}

%% file: 01-introduction.tex
\section{Introduction}\label{sec:intro}

The growing amount of computing elements in HPC systems inherently presents a new bottleneck in terms of the necessary communication for data distribution as well as controlling the application processing elements, e.g., tasks, threads, and processes.
Therefore, the performance of parallel applications highly depends on their communication behavior.

Nowadays, parallel applications execute on a broad range of parallel computing architectures, from large supercomputers to embedded low-power architectures.
When these applications execute on parallel systems, their communication time is affected by how intensely their processing elements exchange data, by the capacity and performance of the network links, and by \emph{the placement of processing elements on the computing resources}.

Application placement is typically the result of a mapping algorithm.
Efficient application placement on modern hardware architectures is of paramount importance for performance~\cite{hoefler2014, ETP4HPC-SRA4:2020, SensiGH:2019, Mahmud:2019, Salaht:2020}.
A poor choice of the mapping algorithm may lead to larger communication latencies and, therefore, to significant performance loss and energy waste.
A plethora of \textit{communication and/or topology-aware mapping} algorithms emerged over the years in the literature to improve application process placement (see \cite{hoefler2014} for a recent overview).




Process mapping, also known as topology mapping or application process placement, is an active area of research with a vast history that includes algorithms whose performance benefits have been recorded in numerous situations.
For instance, communication and/or topology-aware mapping techniques combine information about the target application (its \textit{communication pattern} or \textit{virtual topology}) and the target system (its \textit{physical topology}) to take mapping decisions following a performance objective, e.g., minimizing \emph{dilation}, \emph{distance}, \emph{volume$\cdot$distance}, or congestion~\cite{hoefler2014}.
\JK{Other approaches simply employ \textit{space-filling curves} (SFCs)~\cite{SFC:2013} which essentially generate mappings based on common communication patterns.}

Given the need to achieve high performance and mitigate resource waste, many application developers and users face the following questions:
\begin{enumerate}
	\JK{ \item \textit{How can one verify if an application is suffering from poor communication performance?}}
	\JK{ \item \textit{How much does process mapping impact an application's performance on a given system?}}
	\item \textit{Which mapping algorithm is the highest performing for a given application--system pair?}
\end{enumerate}

The absence of simple answers to these questions has severe consequences.
While a na\"ive process mapping may lead to performance loss, an inapt mapping may cause longer execution times in addition to the overhead associated with generating the mapping itself.
Over time, such performance loss for one or several applications translates into congested or wasted resources and increased energy consumption.
Moreover, performing repeated experiments to identify the highest performing mapping algorithm is neither sustainable nor scalable.
This results in process mapping often being ignored as an explicit application optimization step.


In this work, we study the impact of application process mapping on several processor topologies.
We propose a workflow to support mapping as an explicit application optimization step.
We apply the workflow to four applications, 
mapped using twelve mapping algorithms, 
executing on three direct network topologies. 

This work makes the following contributions:
(i)~it proposes a generic workflow to support process mapping as an explicit application optimization step;
(ii)~it provides an analysis on the predicted mapping benefit for a given application--system pair; and
(iii)~it contributes a Python-based library with well-known topology mapping algorithms from the literature.


This work is organized as follows.
The work related to application optimization through careful placement is reviewed in Section~\ref{sec:rw}.
The generic workflow for mapping applications onto processor topologies is introduced and described in Section~\ref{sec:pa}.
The application characteristics and evaluation metrics are presented in Section~\ref{sec:parApplications}, while the processor topologies and network models are discussed in Section~\ref{sec:topologies}.
Section~\ref{sec:maptomachines} provides a description of the mapping algorithms. 
The design of performance experiments and their evaluation and analysis are presented in Section~\ref{sec:experiments}. Section~\ref{sec:conclusion} concludes the work and outlines future work directions.

%% file: 02-relatedWork.tex
\section{Related Work}\label{sec:rw}

Process placement on modern hardware architectures has been studied from various dimensions and in various contexts, leading to a significant body of work emerging in the literature over the years.
Recent surveys provide an overview of existing solutions and open problems on the topic~\cite{hoefler2014}~\cite{Mahmud:2019}~\cite{Salaht:2020}~\cite{ACMSurveyCommModels:2019}.

The approach taken in this work involves: parallel MPI applications, algorithmic strategies for topology mapping, three-dimensional direct network topologies, communication models, application tracing, and trace-driven simulation.

Most existing work considers parallel MPI applications when studying process mapping.
MPI point-to-point calls have been found in a recent study~\cite{Laguna:2019} to be more prominently used than either persistent point-to-point or one-sided MPI calls. In this work, we study the effect of process mapping on the performance of MPI point-to-point calls.


Hoefler et al.~\cite{hoefler2014} classify the algorithmic strategies for topology mapping into four categories.
The algorithms considered in this work fall into three of those categories: \emph{greedy} includes the \texttt{Peano}, \texttt{Hilbert}, \texttt{Gray}, \texttt{sweep}, and \texttt{scan} SFCs, \texttt{greedy}, \texttt{FGgreedy}, \texttt{greedyALLC}, and \texttt{topo-aware}, \emph{graph partitioning} includes \texttt{bipartition} and \texttt{PaCMap}, while \texttt{Bokhari} is an \emph{isomorphism-based} algorithm.
We do not include any \emph{subgraph isomorphism-based} algorithms, because we only consider bijective mappings in this work.



\FC{Process mapping is influenced by the underlying network topology.
	Therefore, mapping has been studied in the context of modern network topologies (\texttt{torus}, \texttt{fat-tree}, and \texttt{Dragonfly}) and technologies (InfiniBand, Ethernet, BlueGene, Cray, and others)~\cite{hoefler2014}.}\todo{Candidate for removal if short on space.}
~LibTopoMap~\cite{greedy2011} is a generic library of graph mapping heuristics (recursive bisection, $k$-way partitioning, simple greedy strategies, and Cuthill-McKee).
Mapping with LibTopoMap is based on similarity metrics (e.g., bandwidth of the adjacency matrices), employs rank reordering for MPI applications, and can be used on various network topologies and technologies.
While the library is generic and versatile, it is not directly usable in a simulated environment where certain practical effects in the software stack are abstracted and a study of process mapping can concentrate on specific aspects such as the communication cost.
Simulation is also important for the co-design of applications and future systems and is the approach we take in the present work.


MPIPP~\cite{MPIPP:2006} is a framework dedicated to MPI applications with arbitrary virtual topologies executing on SMP clusters and multi-clusters.
Similar to our work, the framework employs application tracing to obtain the communication behavior of the application.
In contrast to this work, the communication pattern is stored as a communication graph which is placed on the topology graph (determined on-the-fly via a parallel ping-pong mechanism).
Also different from our work, they only consider graph partitioning algorithms and conduct direct experiments on SMP clusters.


\FC{Rodrigues et al.~\cite{Rodrigues:2009} use a purely quantitative approach to apply resource binding for MPI applications and reduce communication costs in multicore nodes.
	Mercier and Clet-Ortega~\cite{Mercier:2009} use a similar but qualitative approach to the same problem.
	Both works~\cite{Rodrigues:2009}~\cite{Mercier:2009} employ graph partitioning to compute the mapping.}\todo{This paragraph can be left out if we run low on space as they target multicore architectures, while implicitly only consider multicores... but explicitly consider nodes with distributed memory.}

Automated mapping has also been studied in the context of regular application communication graphs on 2-D and 3-D \texttt{mesh} and \texttt{torus} networks~\cite{hopbyte2010automate}.
There, the virtual topology is also represented as a graph and the mapping heuristics are chosen to optimize for hop$\cdot$Byte.
In our work, the virtual topology is represented as an adjacency matrix, while the mapping heuristics also optimize for  dilation (measured as hop$\cdot$Byte) on 3-D \texttt{mesh} and \texttt{torus} topologies, complemented by the 3-D \texttt{HAEC} \texttt{Box} topology.


EagerMap~\cite{EagerMap:2019} employs a greedy topology mapping algorithm for hierarchical machine topologies (trees).
Its algorithm groups together the application's processes that show the highest affinity based on the communication matrix.
Although the algorithm has been adapted to handle arbitrary network topologies, it requires hierarchical multicore nodes. \FC{In contrast, in this work we concentrate on 3-D network topologies}.

Rico-Gallego et al.~\cite{ACMSurveyCommModels:2019} surveyed prominent communication performance models in high-performance computing, which are often tested in simulation.
They state that future models will need to take into account accurate performance and energy modeling.
In this work, we employ a contention-oblivious communication model called \emph{network coding dynamic resilient} (\ncdr)~\cite{haec_sim_verify} that transmits messages efficiently, reliably, and with minimal energy costs.
\ncdr is implemented in \mbox{HAEC-SIM}~\cite{haec_sim}, the trace-based simulator used in this work.

In addition to careful process mapping, Sensi et al.~\cite{SensiGH:2019} show that application-aware routing outperforms application-agnostic routing.
Similarly, in this work we employ \emph{shortest path routing} and argue that, in addition, \mbox{communication-aware} mapping is needed and show that, in certain cases, it outperforms \mbox{communication-oblivious} mapping.

Trace-driven simulation has also recently been used by Tsuji et~al.~\cite{SCAMP:2019} to study application behavior on future systems.
Their workflow is very similar to ours with the difference that their focus is to support the analysis of applications larger than the real MPI traces from existing systems and they do not explicitly consider process mapping.
While we are also concerned with scalability (planned for future work), in the present work we concentrate on answering the three research questions (Section~\ref{sec:intro}) for modern application-system pairs.



Kenny et al.~\cite{Kenny:2018} consider the influence of the network and its parameters on the performance of parallel MPI applications.
They decompose the application time spent in MPI into: communication, synchronization, and software stack components.
Through the combination of Bayesian inference and trace replay, they found that synchronization and MPI software stack overheads are at least as important as the network itself in determining time spent in communication routines.
In this work, we explicitly study the impact of network topology and its parameters under various process mapping strategies.
Measuring the time spent in synchronization and in various components of the software stack is highly complex and will be incorporated into our workflow in the future.

%% file: 03-proposedApproach.tex
\section{Proposed Workflow}\label{sec:pa}

\begin{figure}[!htb]
	\centering
	\includegraphics[width=0.71\linewidth]{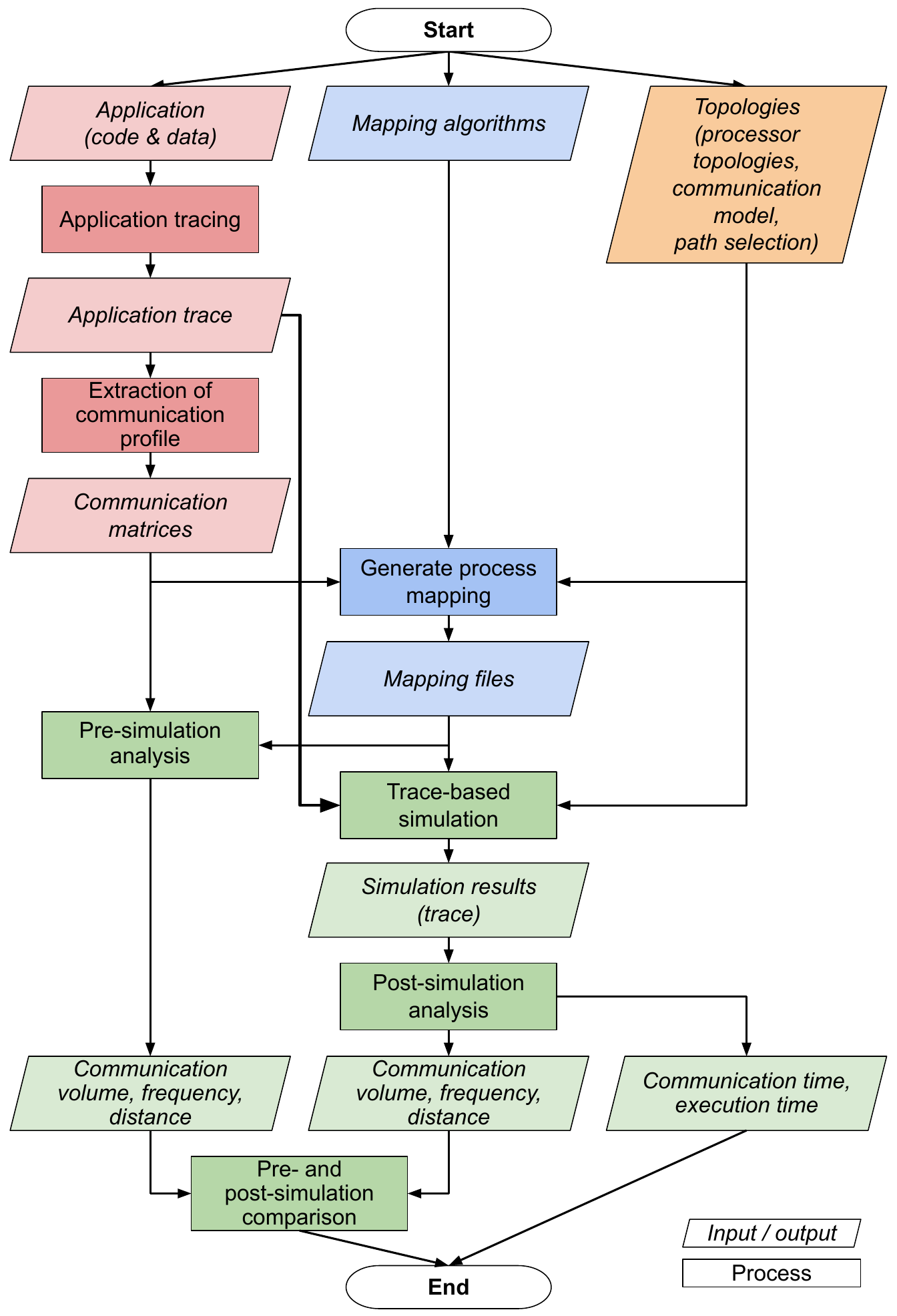}
	\caption{Rectangles represent actions, while parallelograms represent information that actions need as input or produce as output.
		The shape colors represent different types of workflow steps:
		\textcolor{flowred}{red} steps relate to applications;
		\textcolor{flowblue}{blue} steps denote mapping-related activities;
		the \textcolor{floworange}{orange} step concerns machine topologies;
		and \textcolor{flowgreen}{green} steps indicate simulation, performance evaluation, and analysis.}
	\label{fig:workflow}
\end{figure}

This work proposes a workflow, illustrated in Fig.~\ref{fig:workflow}, to render mapping as an explicit application optimization step.
Following the Y-chart design methodology~\cite{Kienhuis:2002}, the workflow provides application optimizers quantitative data obtained by analyzing the performance of applications on topologies for a given set of mapping algorithms.

The steps in \textcolor{flowred}{red} (Section~\ref{sec:parApplications}) refer to the extraction and analysis of performance metrics from the application.
These steps are independent from the mapping strategies or target topologies.
The topology information (in \textcolor{floworange}{orange}, Section~\ref{sec:topologies}) includes the specification of the target topology, communication model, and path selection.
The \textcolor{flowblue}{blue} workflow steps (Section~\ref{sec:maptomachines}) combine the information gathered from applications and topologies to generate various process mappings with mapping algorithms from the literature.

The \textcolor{flowgreen}{green} steps (Section~\ref{sec:experiments}) propose the evaluation and analysis of the outputs from previous steps assessing the performance gains of various mappings for a given application--system pair.
We evaluate and analyze performance in two phases.
The first phase corresponds to analysis \emph{pre-simulation} of the mapping quality using metrics that do not require direct nor simulated experiments.
The second phase corresponds to analysis \emph{post-simulation}, to quantify the application performance gains predicted via simulation.
Finally, we compare the \emph{pre-simulation} results in terms of communication volume, frequency, and distance against those extracted \emph{post-simulation}, to ensure the correctness of the simulation and to assess the usefulness of the pre-simulation metrics.

%% file: 04-applications.tex
\section{Parallel Communication-Intensive Applications}\label{sec:parApplications}
\subsection{Applications and Tracing}
We consider four target applications from well-known benchmark suites.
From the NAS parallel benchmarks~\cite{nasbench}\cite{naswebsite}, we investigated CG and BT-MZ, input size class C~\cite{nassizes}.
NAS CG computes an approximation of the smallest eigenvalue for a large, sparse, and symmetric matrix.
It stresses several irregular long distance communications.
NAS BT-MZ is a newer version of the Block Tri-diagonal solver (BT) which was \mbox{re-designed} to exploit multiple levels of parallelism (MPI and OpenMP).
From the CORAL2~\cite{coral2} benchmark suite, we selected AMG~\cite{amg} with the default input (problem~1).
AMG is a multigrid solver that generates many small messages and stresses memory and network latency.
From the CORAL~\cite{coral} benchmark suite, we used LULESH~\cite{lulesh, luleshpaper} and set the time iterations count to 1,000.
LULESH is a proxy application that simulates a variety of problems that describe the motion of materials relative to each other when subject to forces.


Given the application source codes and input data, we executed and traced them on a parallel system with Intel Broadwell E5-2640 v4~\cite{xeon} processors organized into $2$ sockets, each with $10$ CPU cores.
The nodes in the system are connected by an Intel Omni-Path network with 100 Gbit/s and a two-level \texttt{fat-tree} interconnection topology.
The applications executed on $16$ nodes of this system, using $4$ MPI ranks per node ($2$ on each socket of the node) for a total of $64$ MPI ranks.
To record the application trace, we used the Score-P~v.~4.1~\cite{Knuepfer_2012_ScoreP} measurement infrastructure.

Table~\ref{table:compcomratio} summarizes for each application the time spent in computation versus communication as recorded in the trace.
\FC{The computation time includes all non-MPI functions. }
The MPI total communication time includes time spent in MPI point-to-point operations as well as other MPI calls (MPI~other).


\begin{table}[!htb]
	\caption{Computation and communication profile of the applications.}
	\resizebox{\columnwidth}{!}{%
		\begin{tabular}{l|r|r|r|r|r|r|r|r}
			& \multicolumn{2}{c|}{\textbf{NAS CG}} & \multicolumn{2}{c|}{\textbf{NAS BT-MZ}} & \multicolumn{2}{c|}{\textbf{CORAL2 AMG}} & \multicolumn{2}{c}{\textbf{CORAL LULESH}} \\ \midrule
			
			\textbf{Computation total} & \SI{140.45}{\second} & \SI{2.8}{\percent} & \SI{860.88}{\second} & \SI{84.4}{\percent} & \SI{711.32}{\second} & \SI{75.8}{\percent} & \SI{14231.36}{\second} & \SI{83.2}{\percent} \\ \midrule
			MPI\_Send         & \SI{3628.63}{\second} & \SI{71.3}{\percent} & \multicolumn{1}{c|}{--} & \multicolumn{1}{c|}{--}       & \SI{0.17}{\second} & \SI{0.0}{\percent} & \multicolumn{1}{c|}{--} & \multicolumn{1}{c}{--}          \\
			MPI\_Receive      & \multicolumn{1}{c|}{--} & \multicolumn{1}{c|}{--}    & \multicolumn{1}{c|}{--} & \multicolumn{1}{c|}{--}       & \SI{1.64}{\second} & \SI{0.2}{\percent} & \multicolumn{1}{c|}{--} & \multicolumn{1}{c}{--}          \\
			MPI\_Isend        & \multicolumn{1}{c|}{--}  & \multicolumn{1}{c|}{--}  & \SI{4.06}{\second} & \SI{0.4}{\percent} & \SI{3.13}{\second} & \SI{0.3}{\percent} & \SI{19.74}{\second} & \SI{0.1}{\percent}        \\
			MPI\_Irecv        & \SI{1.71}{\second} & \SI{0.0}{\percent} & \SI{0.53}{\second} & \SI{0.1}{\percent} & \SI{0.41}{\second} & \SI{0.0}{\percent} & \SI{2.29}{\second} & \SI{0.0}{\percent}       \\
			MPI\_Wait         & \SI{1301.72}{\second} & \SI{25.6}{\percent} & \multicolumn{1}{c|}{--} & \multicolumn{1}{c|}{--}       & \multicolumn{1}{c|}{--} & \multicolumn{1}{c|}{--}        & \SI{729.47}{\second} & \SI{4.3}{\percent}       \\
			MPI\_Waitall      & \multicolumn{1}{c|}{--} & \multicolumn{1}{c|}{--}    & \SI{126.97}{\second} & \SI{12.4}{\percent}       & \SI{90.39}{\second} & \SI{9.6}{\percent} & \SI{4.04}{\second} & \SI{0.0}{\percent}        \\
			MPI other        & \SI{13.78}{\second} & \SI{0.3}{\percent} & \SI{27.91}{\second} & \SI{2.7}{\percent} & \SI{130.88}{\second} & \SI{13.9}{\percent} & \SI{2114.95}{\second} & \SI{12.4}{\percent} \\ \midrule
			\textbf{MPI total}         & \SI{4945.84}{\second} & \SI{97.2}{\percent} & \SI{159.47}{\second} & \SI{15.6}{\percent} & \SI{226.62}{\second} & \SI{24.0}{\percent} & \SI{2870.59}{\second} & \SI{16.8}{\percent} \\ \midrule
			\textbf{Total computation $+$ MPI}         & \SI{5086.29}{\second} & \SI{100}{\percent} & \SI{1020.35}{\second} & \SI{100}{\percent} & \SI{937.94}{\second} & \SI{100}{\percent} & \SI{17101.95}{\second} & \SI{100}{\percent} \\ 
		\end{tabular}
	}
	\label{table:compcomratio}
\end{table}



\subsection{Process-logical Communication Matrices}\label{subsection:commMatrices}
The next step involves extracting the application communication behavior from the trace into a \emph{process-logical communication matrix}.
The process-logical communication matrix of the applications can also be collected via other approaches in a comma-separated-value (CSV) format.
The extracted information can represent various quantities such as the number of message exchanges, the volume of exchanges, the average message transfer time, and others.

\begin{figure}[]
	\centering
	\subfigure[CG (exchanges)]{
		\includegraphics[width=.33\linewidth]{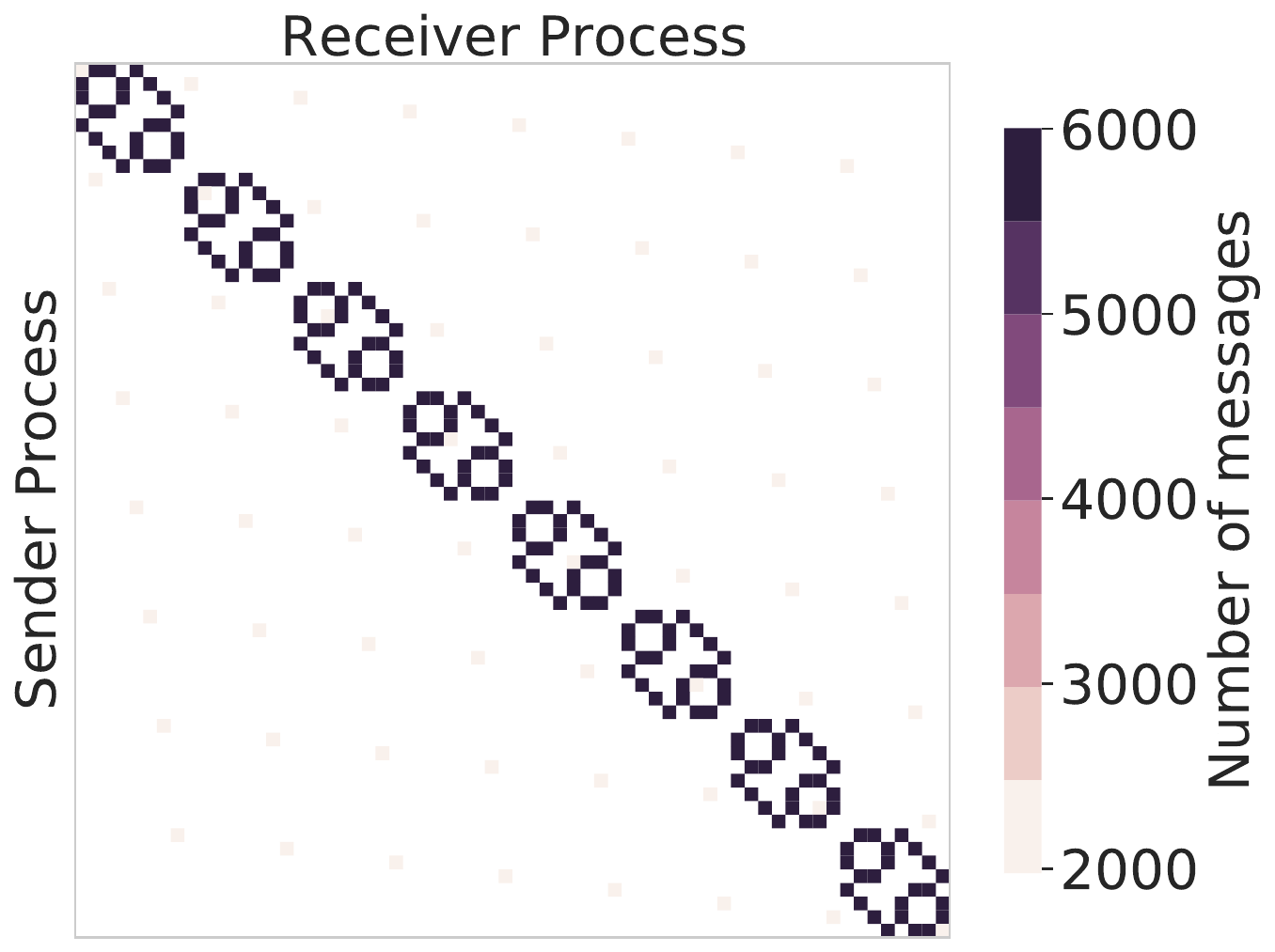}
	}
	\subfigure[BT-MZ (exchanges)]{
		\includegraphics[width=.33\linewidth]{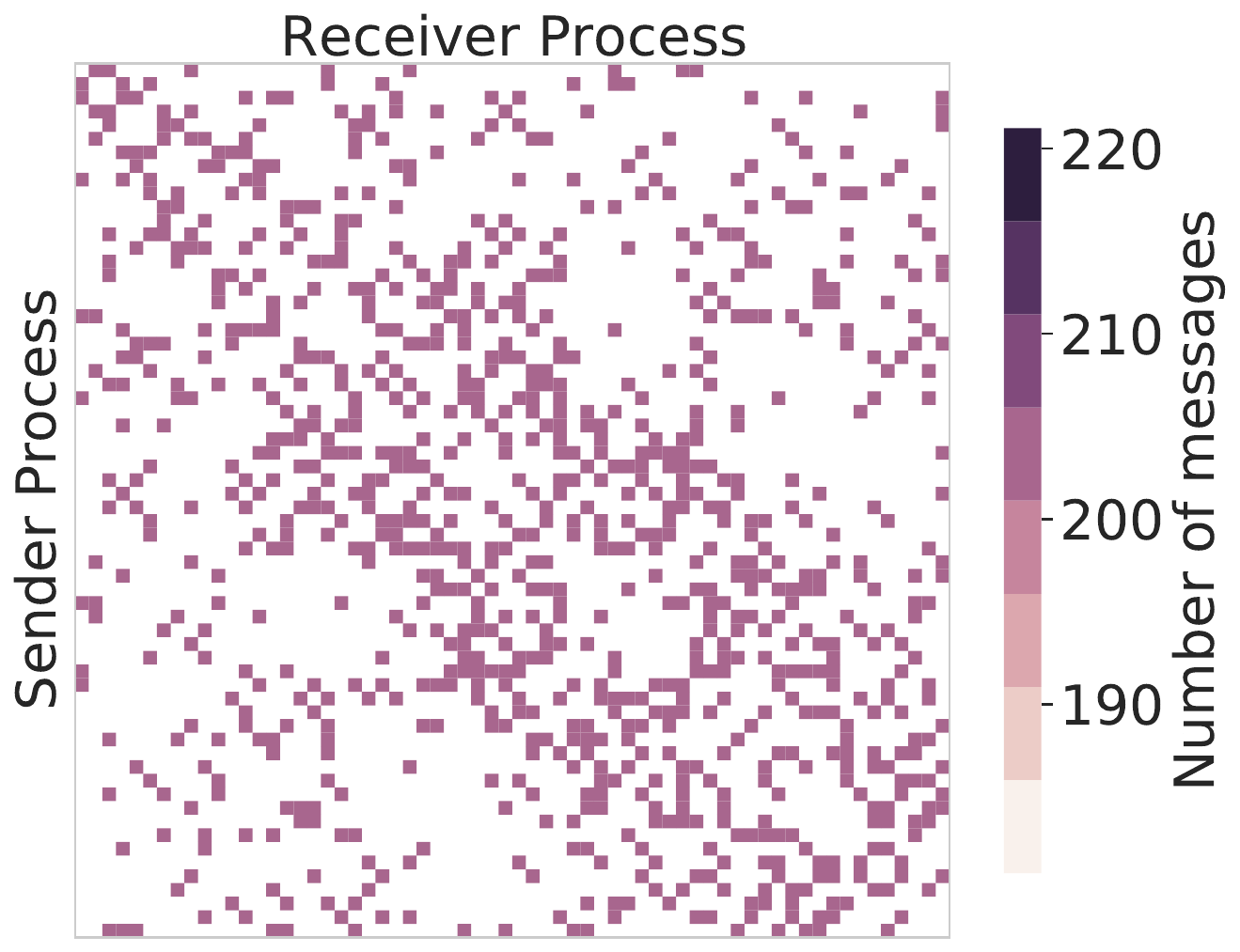}
	}
	\subfigure[LULESH (exchanges)]{
		\includegraphics[width=.33\linewidth]{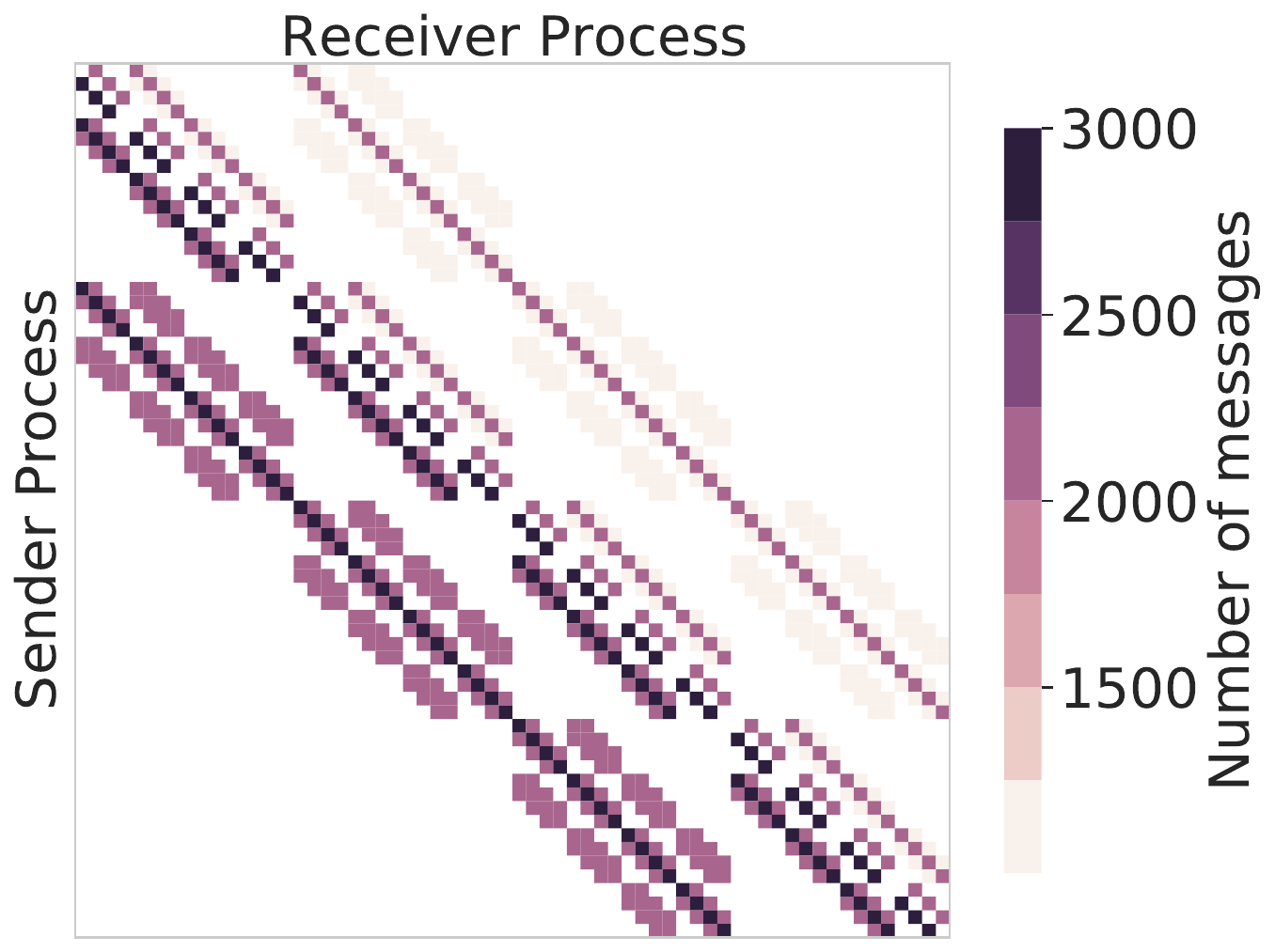}
	}
	\subfigure[AMG (exchanges)]{
		\includegraphics[width=.33\linewidth]{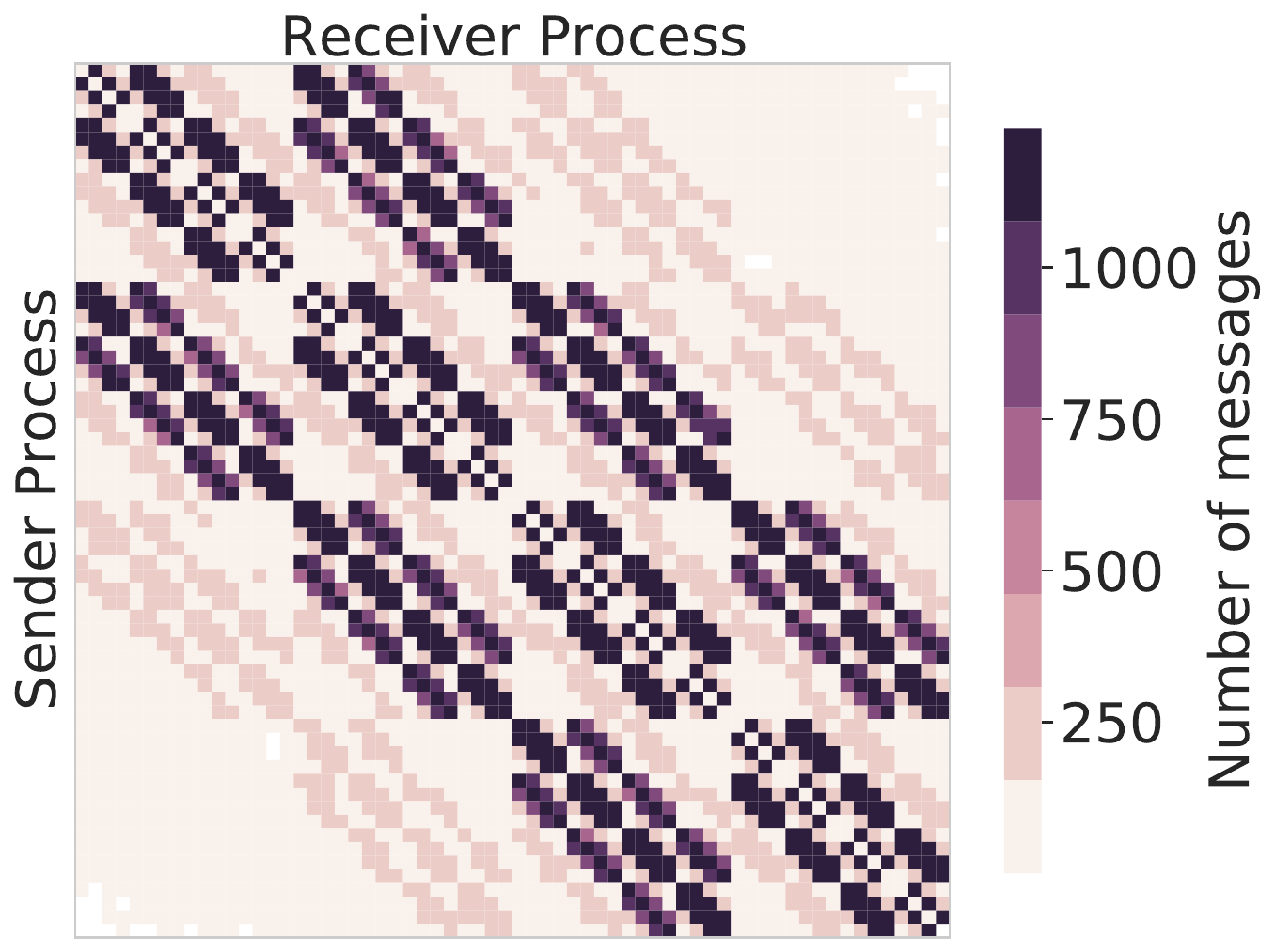}
	}
	\subfigure[CG (volume)]{
		\includegraphics[width=.33\linewidth]{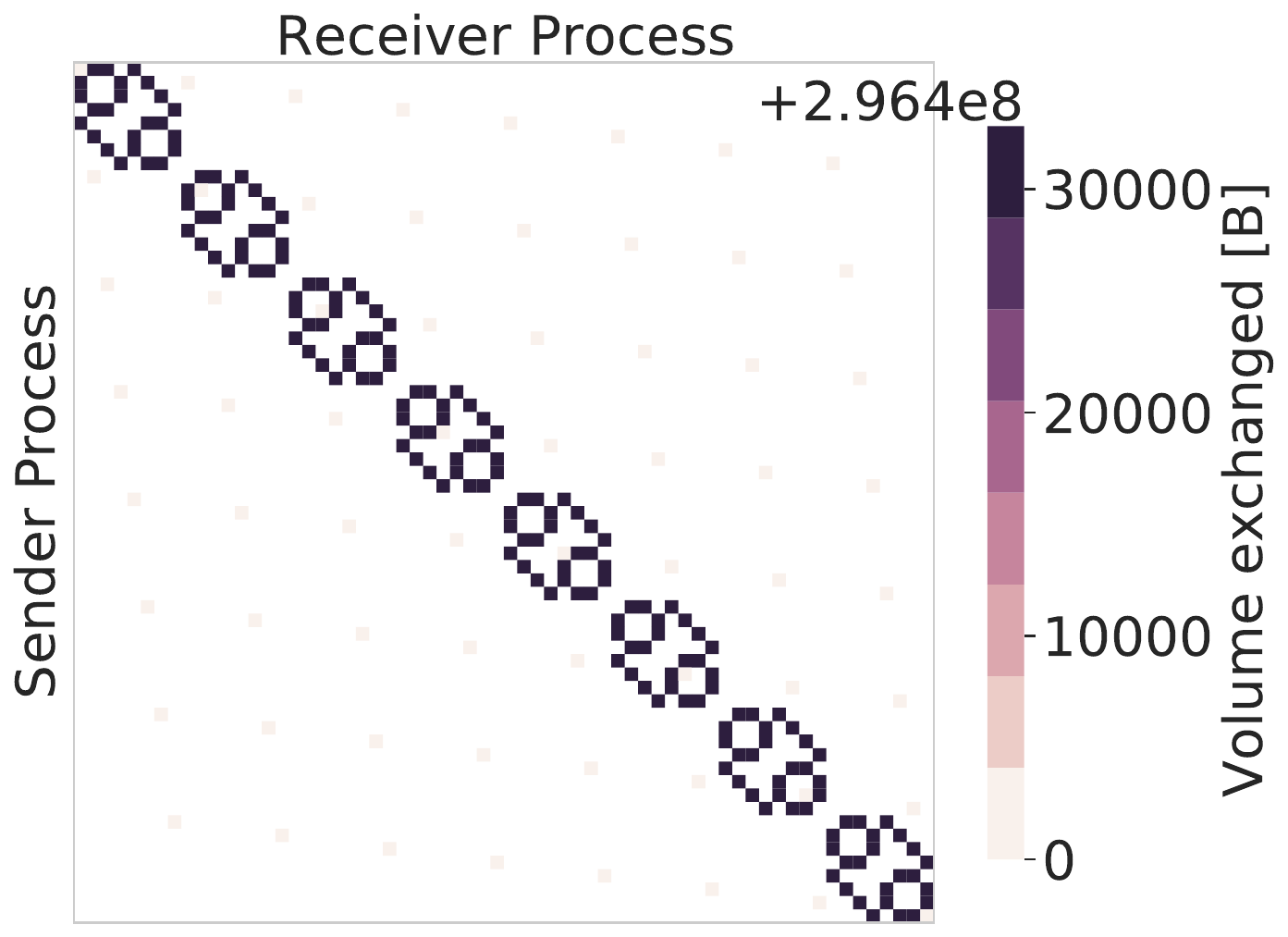}
	}
	\subfigure[BT-MZ (volume)]{
		\includegraphics[width=.33\linewidth]{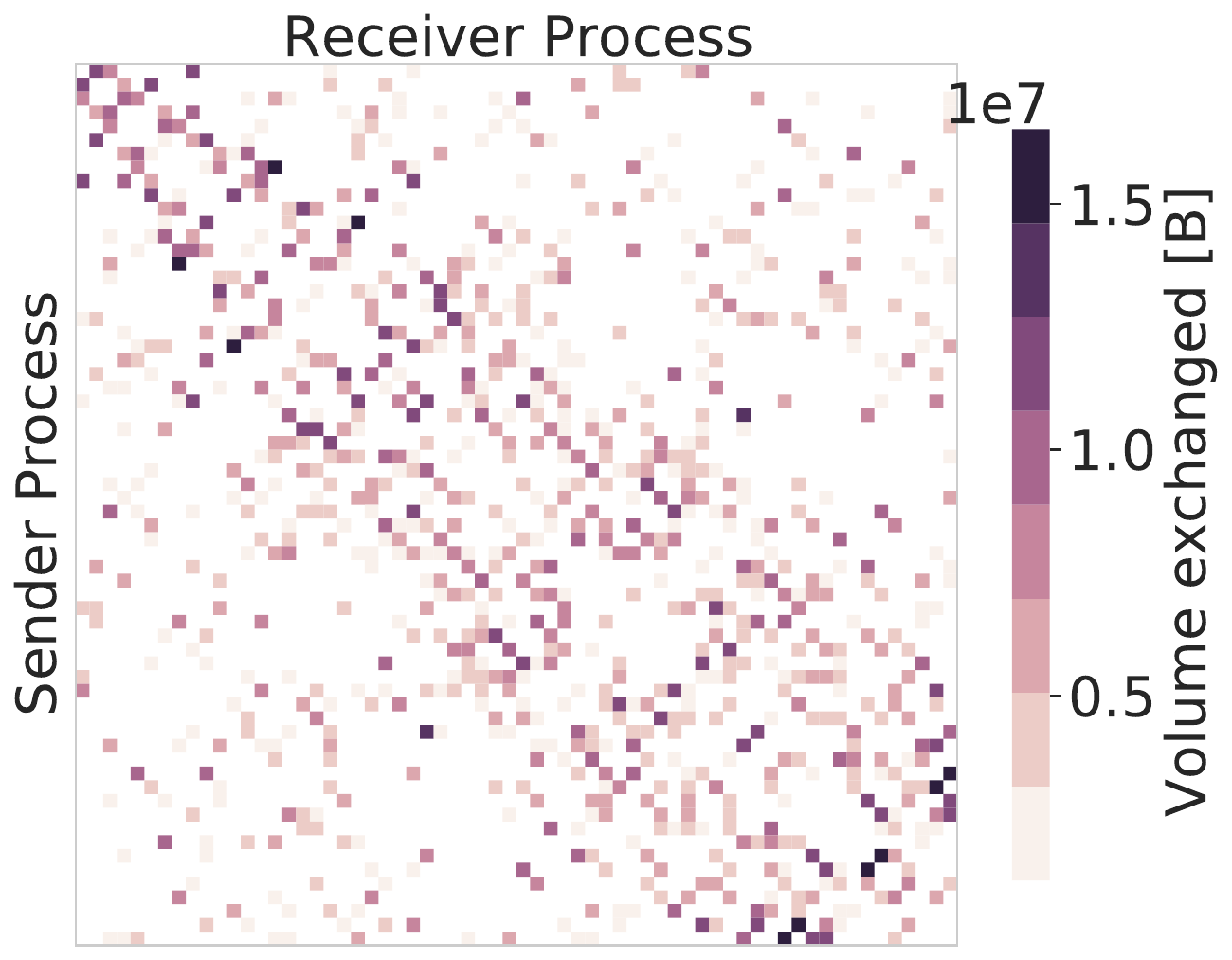}
	}
	\subfigure[LULESH (volume)]{
		\includegraphics[width=.33\linewidth]{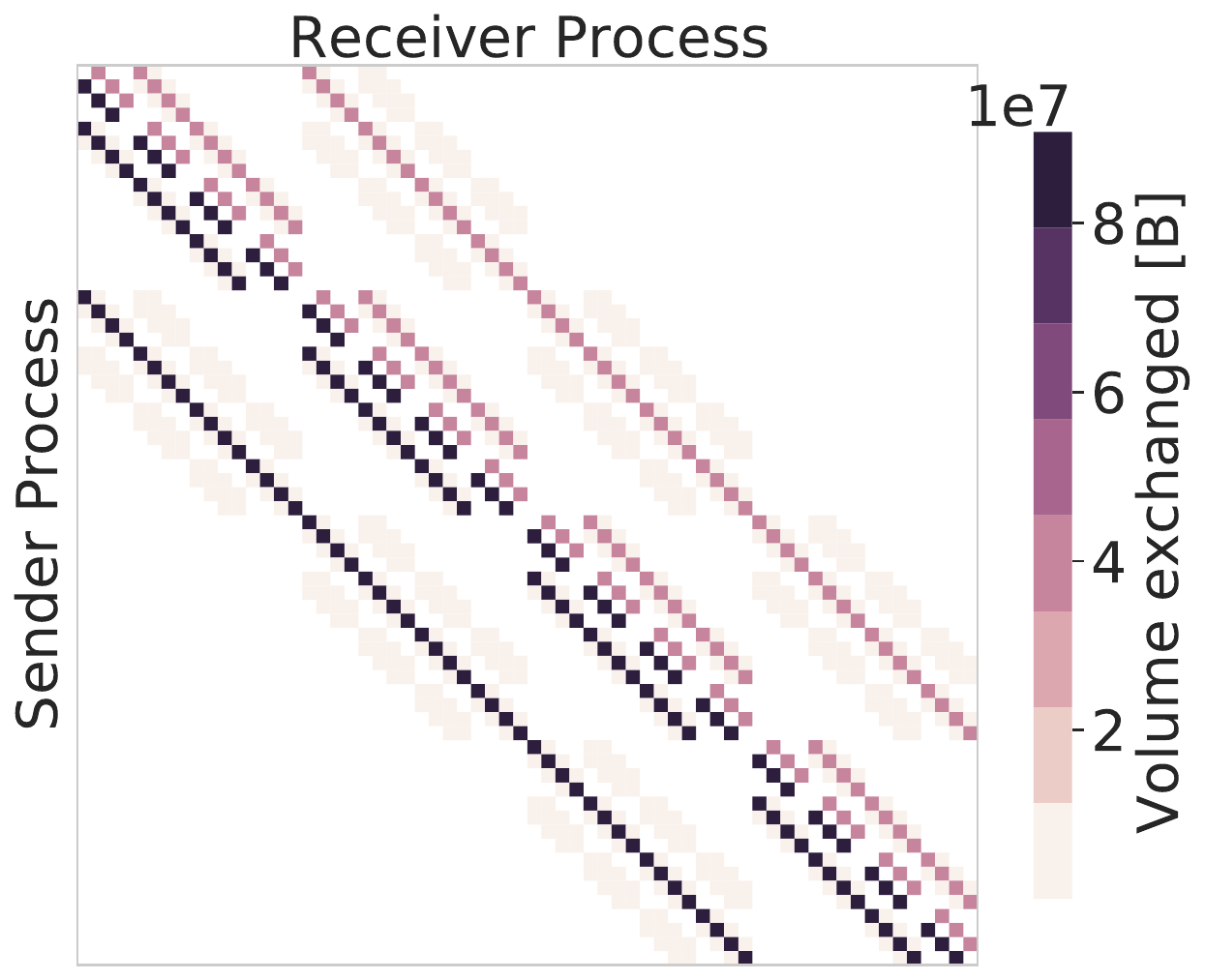}

	}
	\subfigure[AMG (volume)]{
		\includegraphics[width=.33\linewidth]{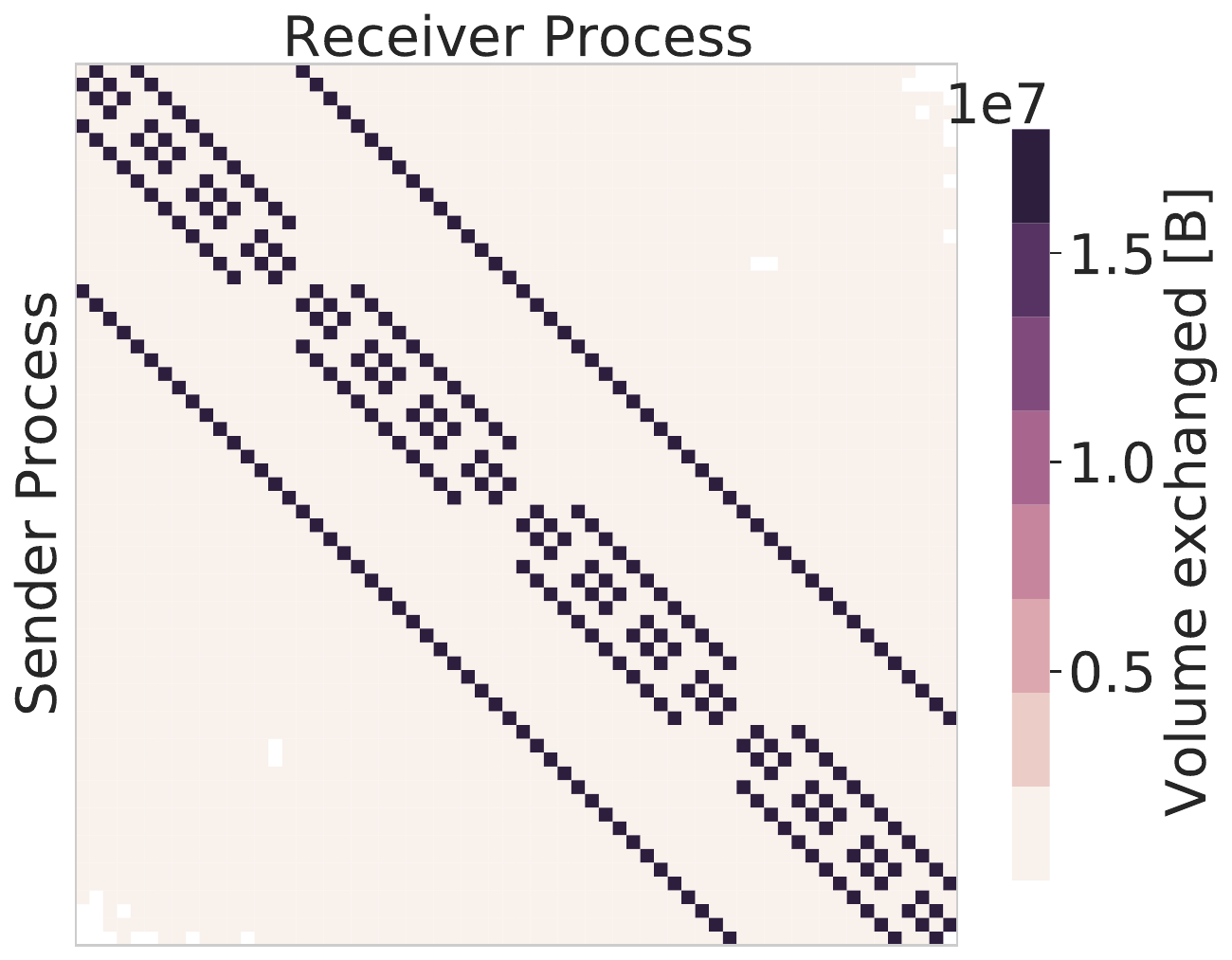}
	}
	\caption[]{Applications' communication behavior for executions with 64 MPI processes.
		The top row shows the communication matrices in terms of point-to-point messages (heat bar: count),
		while the bottom row represents them in terms of volume (heat bar: Byte).
		The $X$ and $Y$ axes represent the receivers and senders, respectively.
		A white cell denotes an absence of communication exchanges.
		The cell in the upper left corner represents exchanges from and to process 0.}
	\label{fig:commMatrices}
\end{figure}

Fig.~\ref{fig:commMatrices} illustrates the two types of process-logical communication matrices employed in this work: number of point-to-point message exchanges (\cmcount) and volume (in Byte) of point-to-point message exchanges (\cmsize).
\JK{BT-MZ, LULESH, and AMG exhibit highly irregular communication patterns which was an important criterion in our applications' selection and already indicates potential room for performance improvement via explicit mapping.}

\subsection{Application-related Communication Metrics}\label{sub:armetrics}

%



Process-logical communication matrices provide a first view into an application's communication behavior.
The structure of a communication matrix alone may prove insufficient to predict if and how much will an application benefit from careful process mapping~\cite{bordage2018}.
To enable such predictions, \textit{communication metrics} (or \textit{matrix-based statistics}) have previously been proposed and tested in the context of thread mapping on shared-memory machines~\cite{diener2015characterizing,diener2015locality} and process mapping on hierarchical machines (multicore nodes in \texttt{fat-tree} topologies)~\cite{bordage2018}.
Such metrics are useful when we are interested in predicting performance gains through mapping, but also to triage applications for which mapping studies can not be performed (e.g., when multiple applications need to be tested while access to resources is very limited) or for which mapping is not expected to yield performance improvements.
\emph{It is also important to emphasize that this work is the first to employ these communication metrics in 3-D topologies}.

The following communication metrics are considered in this work:
\textit{Communication Amount} (\textit{CA}) indicates the average inter-process communication~\cite{diener2015characterizing};
\textit{Communication Balance} (\textit{CB}) measures the divergence of the most communicating process from the  others~\cite{diener2015locality};
\textit{Communication Centrality} (\textit{CC}) quantifies the dispersion of communication from the main diagonal in the communication matrix~\cite{bordage2018};
\textit{Communication Heterogeneity} (\textit{CH}) denotes the average communication variance of each process~\cite{diener2015characterizing};
\textit{Neighbor Communication Fraction} (\textit{NBC}) represents the fraction of communication between processes with close rank identifiers (nearest neighbors)~\cite{bordage2018};
and \textit{Split Fraction} \textit{SP($k$)} measures the fraction of communication among processes in $k^2$ blocks~\cite{bordage2018}.

Depending on the communication matrix used as input (\cmcount or \cmsize) the above metrics represent different values.
For example, \textit{CA} calculated from \cmcount denotes the average number of messages exchanged between all processes, while \textit{CA} calculated from \cmsize indicates the average volume (in Byte) exchanged between all processes.
\emph{For all metrics, higher values indicate a higher potential benefit via careful process mapping.}



We computed the values of the six above metrics for the four applications using the formulae~\cite{commMatrixStatistics}
proposed in~\cite{bordage2018} except for \textit{CA}, which is not mentioned therein and, thus,  we obtained it independently in this work.
We computed the Split Fraction of applications for $k=4$ and~$k=16$ as they represent a portion and a full plane in the 3-D topologies considered, respectively (more details in Section~\ref{sec:topologies}).

Tables~\ref{table:metrics:msg} and~\ref{table:metrics:size} present the values of these metrics based on the \cmcount and \cmsize of the four applications, respectively.
The first row of values presents the sum ($\sum$) of the values of all cells in each communication matrix.
All metric values are rounded upward to three decimal places.
\textbf{Bold} values indicate the largest value for each metric.

\begin{table}[!hbt]
	\centering
	\caption{Communication metrics extracted from \cmcount.}
	\label{table:metrics:msg}
	\resizebox{0.72\columnwidth}{!}{%
		\begin{tabular}{c|cccc}
			& \textbf{CG}	    & \textbf{BT-MZ}		& \textbf{AMG}     	& \textbf{LULESH}  \\
			\midrule
			$\sum$ & \num{1279232} & \num{182910} & \num{1257257} & \num[math-rm=\mathbf]{1692936} \\
			\midrule
			\textit{CA}    	& \num{312.313}     & \num{44.656}         & \num{306.948}        & \num[math-rm=\mathbf]{413.314} \\
			\textit{CB}    	& \num{0.000}       & \num{0.289}          & \num[math-rm=\mathbf]{0.418} & \num{0.413}            \\
			\textit{CC}    	& \num{0.061}       & \num{0.339}          & \num[math-rm=\mathbf]{0.369} & \num{0.325}            \\
			\textit{CH}     	& \num{0.046} 	    & \num[math-rm=\mathbf]{0.171} 	& \num{0.119}		& \num{0.073}            \\
			\textit{NBC}   	& \num{0.700}       & \num[math-rm=\mathbf]{0.963} & \num{0.910}          & \num{0.858}            \\
			\textit{SP(4)} 	& \num{0.387}       & \num[math-rm=\mathbf]{0.941} & \num{0.898}          & \num{0.858}            \\
			\textit{SP(16)} 	& \num{0.074}       & \num[math-rm=\mathbf]{0.651} & \num{0.637}          & \num{0.589}
		\end{tabular}
	}
\end{table} 
\begin{table}[!hbt]
	\centering
	\caption{Communication metrics extracted from \cmsize.}
	\label{table:metrics:size}
	\resizebox{\columnwidth}{!}{%
		\begin{tabular}{c|cccc}
			& \textbf{CG}       & \textbf{BT-MZ} & \textbf{AMG}   & \textbf{LULESH} \\ \midrule
			$\sum$ & \num[math-rm=\mathbf]{75884703744} & \num{4785761760} & \num{5431711224} & \num{20161171008} \\
			\midrule
			\textit{CA}    & \num[math-rm=\mathbf]{18526539.000} & \num{1168398.867}    & \num{1326101.373}    & \num{4922160.809}     \\
			\textit{CB}    & \num{0.000}                & \num{0.205}           & \num[math-rm=\mathbf]{0.273} & \num{0.258}           \\
			\textit{CC}    & \num{0.061}             & \num[math-rm=\mathbf]{0.292} & \num{0.163}          & \num{0.157}           \\
			\textit{CH}    & \num{0.059}             & \num{0.025}          & \num[math-rm=\mathbf]{0.063} & \num{0.041}           \\
			\textit{NBC}   & \num{0.750}             & \num[math-rm=\mathbf]{0.954} & \num{0.686}          & \num{0.677}           \\
			\textit{SP(4)} & \num{0.469}             & \num[math-rm=\mathbf]{0.926} & \num{0.685}          & \num{0.677}           \\
			\textit{SP(16)}& \num{0.187}             & \num[math-rm=\mathbf]{0.584} & \num{0.354}          & \num{0.344}
		\end{tabular}
	}
\end{table} 

Several observations can be drawn from Tables~\ref{table:metrics:msg} and~\ref{table:metrics:size}.
BT-MZ shows the largest values for most metrics, indicating that it has the largest potential to benefit from careful process mapping.
AMG ranks second in terms of the number of highest values. 
Yet, one can note that this occurs for the \textit{CB}, \textit{CC}, and \textit{CH} metrics.
As these metrics are the ones that show the smallest ranges, they may not help in sufficiently differentiating the applications, which makes it difficult to estimate the potential impact of mapping AMG compared to the other applications.

The \textit{CA} metric shows that CG exchanges the largest volumes and LULESH exchanges the highest message counts among all applications.
These two observations indicate that they may be the most sensitive to changes in bandwidth and latency, respectively.
\textit{CB} obtained from both communication matrices are zero for CG.
This means that all processes exchange exactly the same total volume and total number of messages.

These metrics allow us to make predictions of which applications are expected to benefit from careful mapping.
However, the quantification of such benefits is strongly related to the specific network topology used.
For instance, no changes would be observable across mappings on a fully connected topology.
The next section describes the characteristics of network topologies.


%% file: 05-topologies.tex
\section{Interconnection Network Topologies}\label{sec:topologies}
One of the three inputs to the workflow illustrated in Fig.~\ref{fig:workflow} is the target topology (in \textcolor{floworange}{orange}).
The mapping problem might only be apparent when thinking of parallel applications running on different nodes of an HPC system.
However, from an abstract perspective, a self-similar fractal-like observation \LP{unfolds}:

Most modern networks can be seen as connected to an encompassing interconnection network or as an interconnection network to their sub-networks.
For instance, an HPC system comprises several isles, a node contains multiple NUMA domains, and a SoC processor architecture contains several components.
Each of the above systems employs a network topology containing another sub-network.
Hence, there are numerous opportunities to analyze the impact of different mapping strategies on the performance of applications executing in such systems.

\subsection{Direct 3-D Topologies}

Nowadays, HPC systems are usually built with sophisticated network topologies, e.g., \texttt{butterfly}, \texttt{fat-tree}, \texttt{Dragonfly}.
Yet in practice, not every application receives an allocation that allows it to exclusively utilize the entire network.
Instead, a resource allocation forms a virtual subnetwork, which has a simpler topology and requires local process mapping.
Furthermore, other networks still employ simpler topologies.
For instance, the Intel SkyLake SP architecture uses a 2-D \texttt{torus} topology to interconnect all cores on a chip~\cite{schone2019energy}.

Influenced by these observations, we consider topologies that arrange 64 nodes in a $4 \times 4 \times 4 $ 3-D fashion as a proxy for more complex interconnection topologies.
Such an arrangement already allows a rich selection of network topologies, namely 3-D \texttt{mesh}, 3-D \texttt{torus}, and the 3-D \texttt{HAEC} \texttt{Box} topology.

The \texttt{HAEC Box} topology stems out of the HAEC project~\cite{haec_box} which envisions a novel highly adaptive and energy-efficient network topology, with nodes arranged on four vertically laid out (in the $Z$ dimension) boards and each board contains $4 \times 4$ nodes (in the $XY$ plane).
Nodes of one board are interconnected using fast optical links in a \mbox{2-D} \texttt{torus}.
A fully-connected wireless interconnection array facilitates inter-board communication (in the $Z$ dimension).

We assume that all topologies considered in this work employ future-generation network links.
Table~\ref{tbl:links} shows the characteristics of such future-generation links.
While the \texttt{mesh} and \texttt{torus} topologies only contain optical links, the \texttt{HAEC Box} is a \emph{heterogeneous} network topology and employs both wireless (across boards) and optical (on-board) links.
The heterogeneity of the \texttt{HAEC Box} links \FC{mandates a systematic analysis of the suitability of mapping algorithms, such as the one presented in this work.}

\begin{table}[t]
	\centering
	\caption{Network link characteristics used in simulation.}
	\resizebox{0.75\columnwidth}{!}{%
		\begin{tabular}{l|c|c|c}
			\textbf{Link type} & \textbf{Bandwidth}   & \textbf{Latency} & \textbf{Bit error rate} \\
			\midrule
			Wireless  & \SI{100}{\giga\bit\per\second} & \SI{100}{\pico\second}  & \num{1e-8}           \\
			Optical   & \SI{250}{\giga\bit\per\second} & \SI{10}{\pico\second}   & \num{1e-12}
		\end{tabular}
	}
	\label{tbl:links}
\end{table}

\subsection{Path Selection}

In the pre- and post-simulation analyses (Sections~\ref{sub:presim} and~\ref{sub:postsim}), we exclusively use static path selection, i.e., the path of individual packets through the network only depends on the source and destination positions and remains unchanged during execution.
The 3-D \texttt{mesh} and 3-D \texttt{torus} topologies employ \emph{shortest-path $XYZ$ dimension order routing} (DOR) to predetermine the path that the messages will take.
$XYZ$ DOR first routes messages along the $X$ dimension until the $X$ coordinate of the current hop destination is equivalent to the $X$ coordinate of the message destination.
The hop selection is applied similarly along the $Y$ and then the $Z$ dimensions.

On the \texttt{HAEC Box} topology, the simulation uses the same $XYZ$ DOR algorithm when the message source and destination are on the same board.
When the message has to hop across boards, the first hop is to the node on the neighboring board with the same $X$ and $Y$ coordinates as the destination node.
After the first hop across boards, every subsequent hop only has to follow along the $Z$ dimension until the message reaches its destination.
\vspace{-0.3cm}
\subsection{Communication Model}

To model communication, we use a contention-oblivious implementation of the \ncdr network model~\cite{dncr_model}, implemented in \mbox{HAEC-SIM}~\cite{haec_sim}.
This model describes the separation of message bodies into packets and the duration of efficient, reliable, minimal energy cost transmissions over the network based on future-generation wireless and optical links.

%% file: 06-core.tex

\section{Mapping Applications onto Parallel Machines}\label{sec:maptomachines}

\FC{This section describes the \textcolor{flowblue}{blue} workflow steps (Fig.~\ref{fig:workflow}) and the mapping algorithms and clarifies their implementation and use in this work.}
\subsection{\maplib: a Mapping Algorithms Library}
\JK{We implemented twelve mapping algorithms from the literature (see Section~\ref{sec:rw}) in a new Python library called \maplib~\cite{maplib}.}
The algorithms generate mappings for the three 3-D topologies considered herein: \texttt{mesh}, \texttt{torus}, and \texttt{HAEC Box}.
The support for other topologies is considered as future work.
Additional details regarding the installation and particular configurations are offered together with the library.
The library generates ASCII mapping files, the structure of which is exemplified in~\cite{haec_sim}.

\maplib implements:
(i)~Communication- and topology-oblivious mapping strategies (Section~\ref{sub:commob}), which do not take into account the communication matrices of the applications nor the target processor topologies.
These algorithms follow a predetermined node ordering to map all processes to the available nodes and produce deterministic mappings.
(ii)~Communication- and topology-aware mapping strategies (Section~\ref{sub:commtopoaw}), which consider both  communication matrices of the applications and target processor topologies.
These algorithms produce different mappings for a given application--system pair.



\subsection{Communication- and Topology-Oblivious Mapping}\label{sub:commob}
\begin{figure}[!htb]
	\centering
	\subfigure[Peano]{
		\includegraphics[width=.22\linewidth]{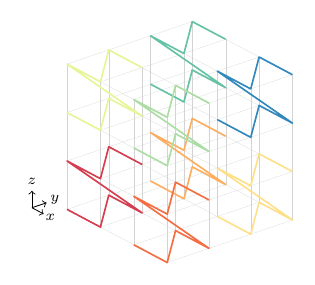}
		\label{fig:peano}
	}
	\subfigure[Hilbert]{
		\includegraphics[width=.22\linewidth]{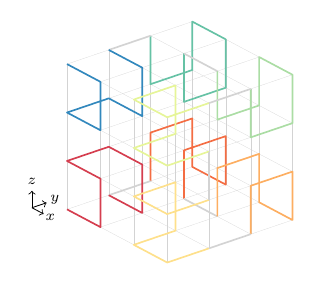}
		\label{fig:hilbert}
	}
	\subfigure[Gray]{
		\includegraphics[width=.22\linewidth]{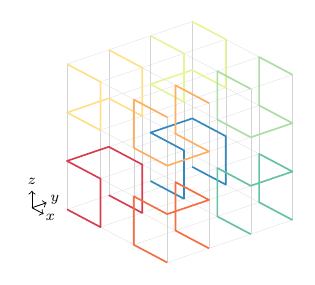}
		\label{fig:gray}
	}
	\subfigure[sweep]{
		\includegraphics[width=.22\linewidth]{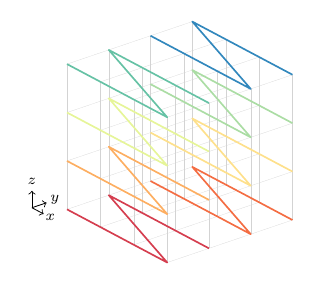}
		\label{fig:sweep}
	}
	\subfigure[scan]{
		\includegraphics[width=.22\linewidth]{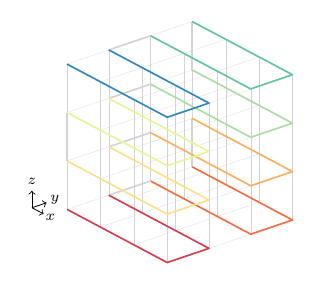}
		\label{fig:scan}
	}
	\caption[]{Illustration of the five SFCs on a 3-D \texttt{mesh} topology. The curves start on the bottom left corner of the topology and proceed along the lines in the \textcolor{red}{red}-\textcolor{orange}{orange}-\textcolor{yellow}{yellow}-\textcolor{olive}{olive}-\textcolor{green}{green}-\textcolor{blue}{blue} order.}
	\label{fig:sfcs}
\end{figure}

The communication- and topology-oblivious mapping strategies implemented are five space filling curves (SFCs), illustrated in Fig.~\ref{fig:sfcs}, that are extensively used as mapping schemes. They map discrete multi-dimensional spaces onto one-dimensional spaces~\cite{mokbel2003analysisSFC}.
SFCs were discovered in the nineteen century by Peano~\cite{peano1890}, followed by Hilbert~\cite{Hilbertber1891}. Numerous SFC variations have been studied since then.

The \texttt{sweep} SFC mapping algorithm is the \emph{straightforward} mapping that follows the nodes in the topology in a $XYZ$ order.
As \emph{default reference mapping} we use \texttt{sweep} to allow the evaluation of the quality and performance improvements generated by all other mappings in \maplib (Section~\ref{sec:experiments}).

\subsection{Communication- and Topology-Aware Mapping}\label{sub:commtopoaw}

\texttt{Bokhari}~\cite{bokhari1981} is a mapping algorithm that originates in graph theory and was initially proposed for the assignment of parallel applications solving structural problems on the finite element machine.
It takes an initial mapping and proceeds in two steps.
In the first step, it searches for the pairwise task interchanges that maximize the cardinality (matching of edges of the application graph to edges in the machine topology) to generate a new mapping.
In the second step, it evaluates if the new mapping has a higher cardinality than the previous one.
If this is the case, it stores the best mapping, generates a new mapping through random task swaps, and returns to the first step.
Otherwise, it returns the best mapping found previously.

The \texttt{topo-aware}~\cite{topoAware2006} algorithm splits mapping into two phases.
The first is a partitioning phase which groups heavily communicating processes in the same task.
The second is the mapping phase in which tasks are mapped onto the processors such that more heavily communicating tasks are placed on nearby processors.
The algorithm uses an estimation function that calculates the cost of placing an unallocated task on an available processor in each cycle.

We implemented three greedy algorithms: \texttt{greedy}~\cite{greedy2011}, \texttt{FHgreedy}~\cite{FHGreedy2015}, and \texttt{greedyALLC}~\cite{greedyALLC2015}.
Both \texttt{greedy} and \texttt{FHgreedy} start mapping the most communicating process to a random node, while
\texttt{greedyALLC} maps the most communicating process to the most connected node.
Their main differences appear in the next steps they perform.
\texttt{greedy} maps the heaviest communicating processes close to each other.
\texttt{FHgreedy} maps neighbors according to the amount of communication that they have with each other.
Lastly, \texttt{greedyALLC} first pairs the most communicating processes then maps them close to each other.

%
%

The \texttt{bipartition} mapping was proposed to improve inter-node mapping on \texttt{torus} and \texttt{mesh} topologies through a recursive bipartitioning algorithm~\cite{bipartitioning22015}.
The mapping is obtained by recursively dividing the application communication graph using a multilevel $k$-way partitioning algorithm~\cite{karypis1998multilevelk} ($k=2$), while the machine topology is simply recursively split in the middle of its largest dimension.

\texttt{PaCMap}~\cite{pacmap2015} is a graph-based algorithm that simultaneously conducts job allocation and process mapping to reduce communication overhead.
This work implements the process mapping step of \texttt{PaCMap}.
It starts by partitioning the application communication matrix into process groups (PGs) of highly-communicating processes.
After that, the algorithm selects a center PG (in this work, a single process) and maps it to a center node in the topology.
Then, it expands the allocation by picking a node and mapping a process to it based on the network topology and the communication graph until all tasks are mapped.
This results in highly-communicating PGs being mapped close to one another.

%% file: 07-experiments.tex
\section{Performance Evaluation}\label{sec:experiments}

In the proposed workflow, performance evaluation consists of three steps (\textcolor{flowgreen}{green} steps in Fig.~\ref{fig:workflow}).
(i)~The \emph{pre-simulation performance analysis} step evaluates the mappings' quality using metrics derived without execution or simulation.
(ii)~The \emph{post-simulation performance analysis} step evaluates the performance of the mappings using simulations.
(iii)~The \emph{pre- and post-simulation performance comparison} shows the impact of mapping and allows the assertion of invariant properties (communication volume and exchanges) post-simulation.

Table~\ref{tbl:parameters} shows the parameters used in the design of the factorial experiments.
The details and findings from these experiments are described in the following.

\begin{table}[!htb]
	\caption{Parameters used in the Design of Factorial Experiments}
	\centering
	\resizebox{0.8\columnwidth}{!}{
		\begin{tabular}{c|l|c|c}
			\multicolumn{2}{c|}{\textbf{Parameter}}    & \multicolumn{1}{c|}{\textbf{Values}} & \textbf{Count}\\ \midrule
			\multicolumn{2}{c|}{\multirow{2}{*}{Application}}   & NAS CG, NAS BT-MZ & 4     \\ 
			\multicolumn{2}{c|}{}  & CORAL2 AMG, CORAL LULESH &    \\ \midrule
			\multirow{1}{*}{Mapping} & \begin{tabular}[c]{@{}l@{}}Communication- \&\\ topology-oblivious\end{tabular}  & \texttt{Peano}, \texttt{Hilbert}, \texttt{Gray}, \texttt{sweep}, \texttt{scan} & 5  \\ \cline{2-4}
			\multirow{1}{*}{algorithm}	& \begin{tabular}[c]{@{}l@{}}Communication- \&\\ topology-aware\end{tabular}  & \begin{tabular}[c]{@{}l@{}}\texttt{Bokhari}, \texttt{topo-aware}, \texttt{greedy}, \texttt{FHGreedy}, \\\texttt{greedyALLC},  \texttt{bipartition}, \texttt{PaCMap}\end{tabular} & 7 \\ \midrule
			\multicolumn{2}{c|}{Mapping input}  & \cmcount, \cmsize & 2 \\ \midrule
			\multicolumn{2}{c|}{3-D topology}  & \texttt{mesh}, \texttt{torus}, \texttt{HAEC~Box} & 3 \\ \midrule
			\multicolumn{2}{c|}{Communication model}  & \ncdr & 1 \\ \midrule
			\multicolumn{3}{c|}{\textbf{Total factorial experiments}} &  \multicolumn{1}{c}{\textbf{288}}  \\
		\end{tabular}
	}
	\label{tbl:parameters}
\end{table}

\subsection{Pre-simulation Mapping Performance Analysis} \label{sub:presim}

The \emph{pre-simulation analysis} step evaluates the mappings' quality using metrics derived without execution or simulation.
This step requires the following information: the application communication behavior, denoted by \cmcount and \cmsize, the process mappings, the specification of the target network topology, and the routing algorithm.

Given this information, one can use metrics such as dilation, average and total number of hops traveled by the application messages, and volume communication transmitted through the network links~\cite{hopbyte06topomap}.

In this work, we use \emph{dilation} as a pre-simulation metric to evaluate the mappings' quality.
Dilation is widely used as a performance and quality metric in process mapping~\cite{topoAware2006,  hopbyte06topomap, hopbyte2013Predicting, bordage2018}, where it is also referred to as `hop-Bytes'.
Dilation is easy to determine and tends to correlate well with the applications' performance~\cite{hopbyte2010automate}.
Dilation, denoted as $D$, can be calculated using (\ref{eqn:hopbyte}), where $P$ denotes the set of application processes, $\delta$ denotes the mapping function, $d$ represents the distance between nodes (number of hops), and $w$ denotes the weight function (e.g., the communication volume in Byte).
\emph{A lower dilation value leads to improved performance and implicitly to a communication energy consumption reduction.}

\begin{equation}
\text{D} = \sum_{i \in  P} \sum_{j \in  P} d(\delta(i), \delta(j)) \cdot w(i,j).
\label{eqn:hopbyte}
\end{equation}

\begin{figure*}[!htb]
	\centering
	\subfigure[CG]{
		\includegraphics[width=.47\linewidth]{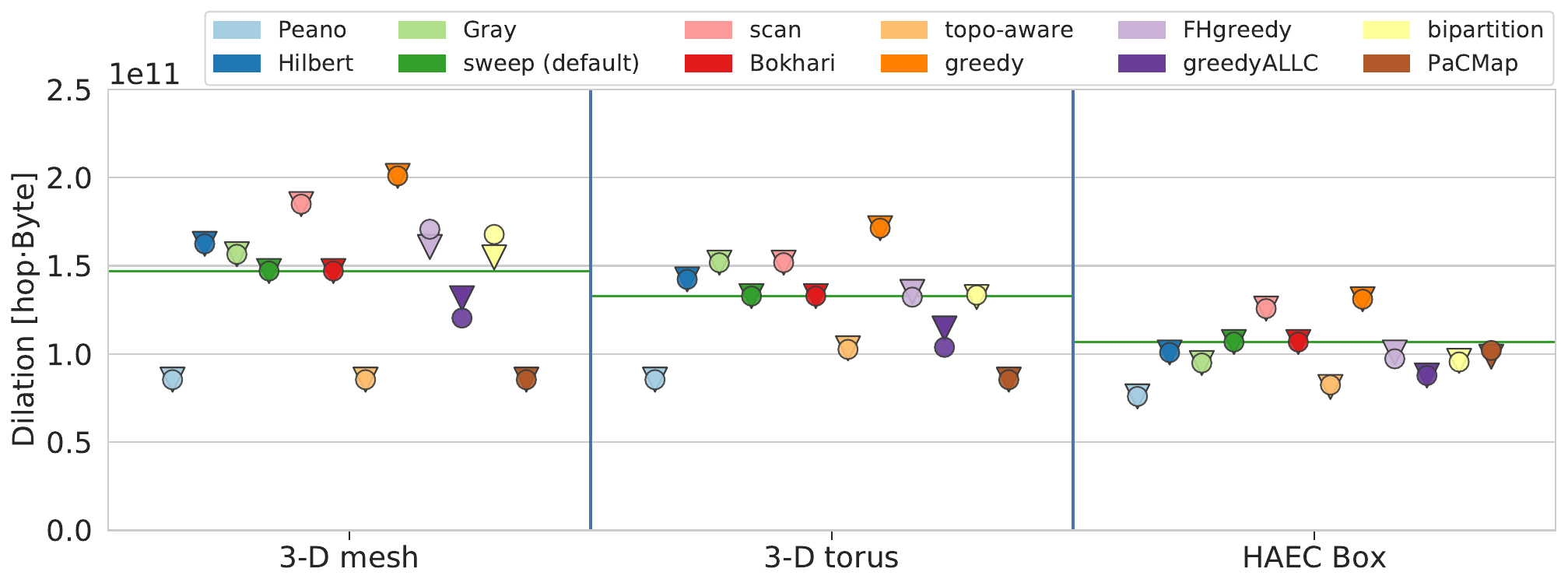}
		\label{subfig:cghop}
	}
	\subfigure[BT-MZ]{
		\includegraphics[width=.47\linewidth]{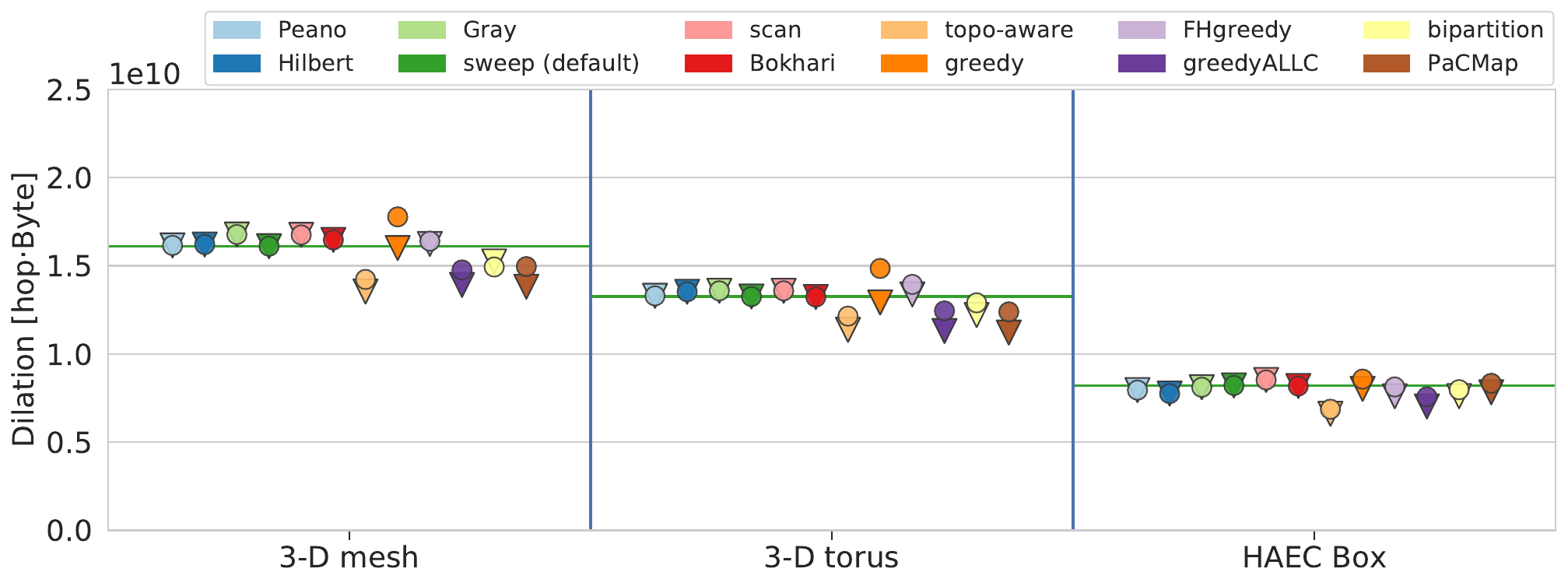}
		\label{subfig:bthop}
	}
	\subfigure[LULESH]{
		\includegraphics[width=.47\linewidth]{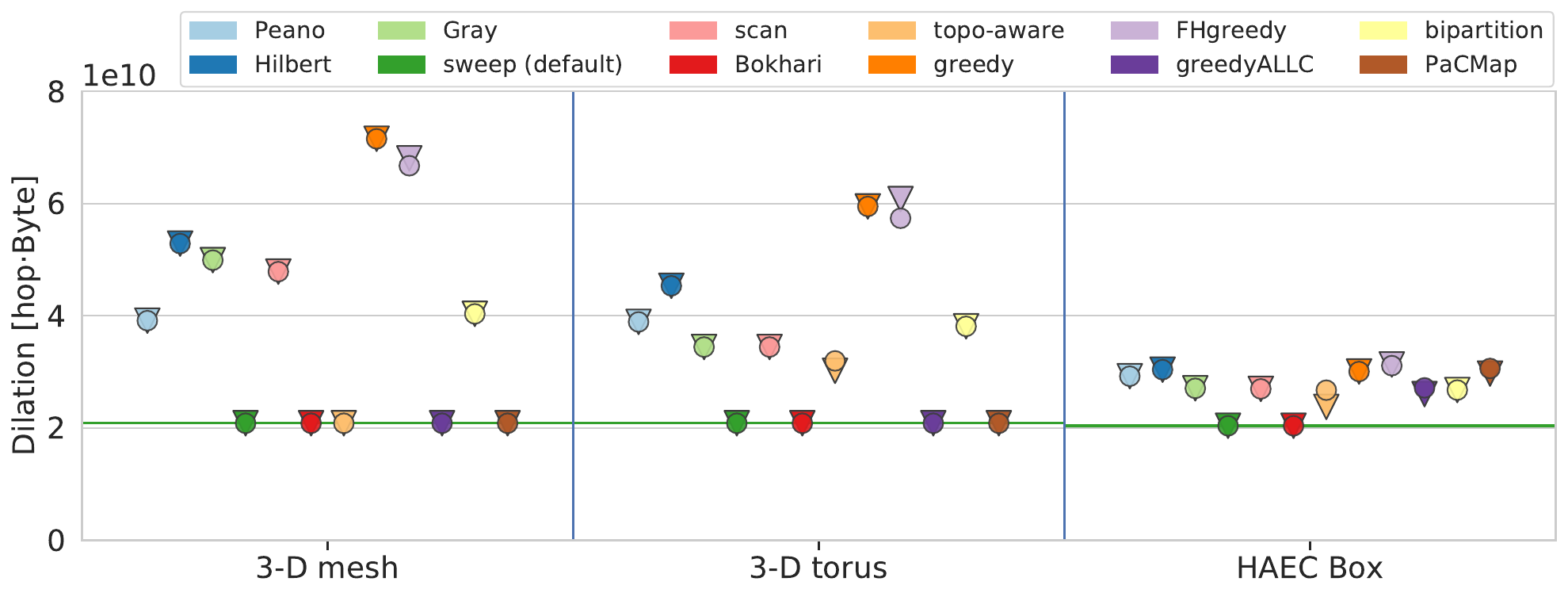}
		\label{subfig:luleshhop}
	}
	\subfigure[AMG]{
		\includegraphics[width=.47\linewidth]{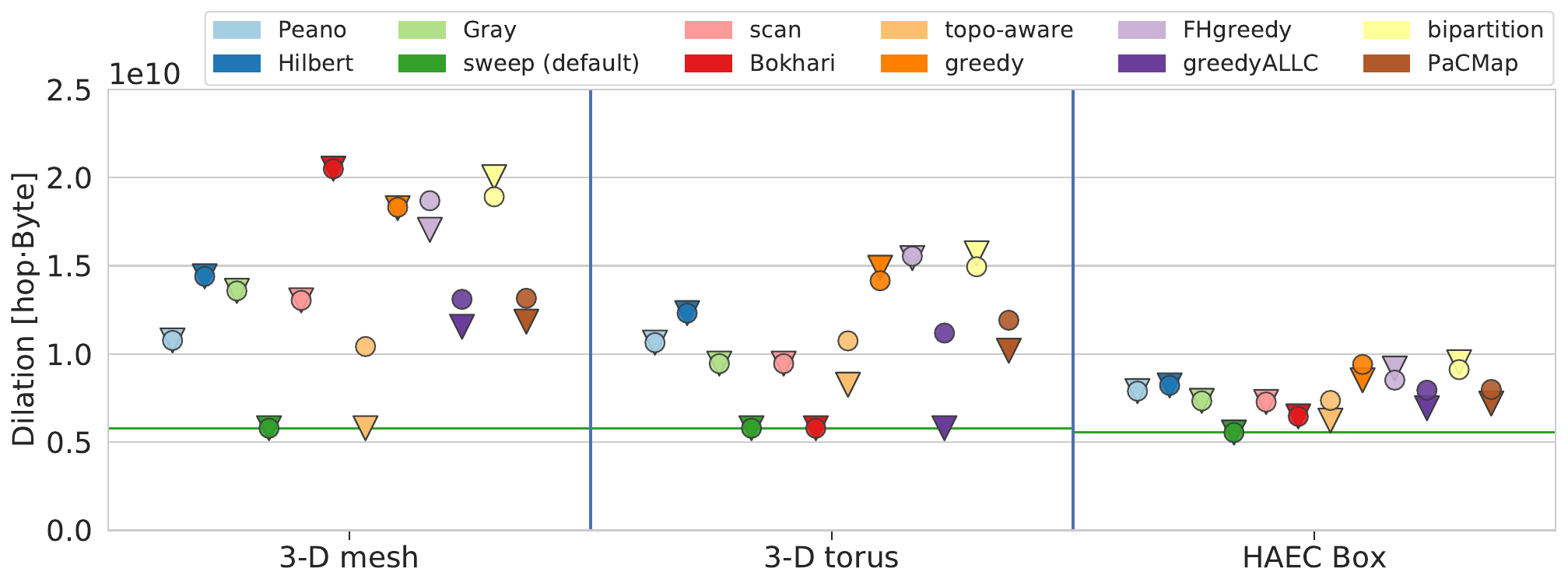}
		\label{subfig:amghop}
	}
	\caption[]{
		Dilation for all applications, mappings, and topologies.
		Topologies are shown on the $X$ axis.
		The dilation values are shown on the $Y$ axis.
		Each mapping is represented with a distinct color.
		The \textcolor{sweepgreen}{green} horizontal line denotes the dilation of the \textcolor{sweepgreen}{\texttt{sweep}} mapping, which we consider as the default mapping on a given topology.
		Circles $\circ$ denote mappings based on \cmcount, while triangles $\triangledown$ indicate mappings based on \cmsize.}
	\label{fig:hopbytes}
\end{figure*}

\JK{Fig.~\ref{fig:hopbytes} shows the dilation associated with each mapping for every application--topology pair.
	%
	One can note
	that dilation significantly varies across topologies and mappings.
	Specifically, most mappings improved performance over \texttt{sweep} (\emph{default}) for CG and BT-MZ.}

CG achieves the lowest dilation with \texttt{Peano}, \texttt{topo-aware}, and \texttt{PaCMap} for 3-D \texttt{mesh}, with  \texttt{Peano} and \texttt{PaCMap} for 3-D \texttt{torus}, and with \texttt{Peano} for \texttt{HAEC} \texttt{Box}.
BT-MZ achieves the lowest dilation with \texttt{topo-aware} for 3-D \texttt{mesh} and \texttt{HAEC} \texttt{Box}.
For \texttt{torus}, the \texttt{PaCMap} mapping with \cmsize achieved the lowest dilation.
These observations indicate that CG and BT-MZ exhibit the potential for performance improvement with these mappings.
LULESH and AMG achieved the lowest dilation with \texttt{sweep} (default) which indicates that they will not benefit from any of the other mappings on these topologies.

\LP{
	These results also illustrate the sensitivity of the performance of mapping algorithms to the application and topology.
	For instance, \texttt{Bokhari} shows the highest dilation for AMG on the 3-D \texttt{mesh} but one of lowest ones on the 3-D \texttt{torus}.
}

Even though the dilation of the mappings performed using \cmcount and \cmsize is comparable, \cmsize helped mappings achieve a slightly lower dilation in most cases.
\emph{This indicates that \cmsize is a mapping input that has a greater impact on performance.}

Directly comparing a mapping algorithm across the topologies, the performance on the \texttt{HAEC} \texttt{Box} topology always yields the lowest dilation due to its increased link connectivity which offers shorter paths for messages to travel on.

\subsection{Application Performance Prediction Using Simulation}\label{sub:simulation}

We use HAEC-SIM~\cite{haec_sim}\cite{HAECSimlinkrepo} to simulate the applications' performance.
HAEC-SIM uses an application's trace as an application model during the simulation process.
With the respective models for the target topology, the simulator generates new traces representing the applications' predicted behavior on the target system.
The simulator works deterministically, i.e., simulations with the same input will repeatedly generate the same output.
HAEC-SIM's parameters are the application trace, the mapping, and the configuration file specifying the target system~\cite{arxivMapping2020}.

While the duration of the computation operations are fixed in the input trace, we use a contention-oblivious implementation of the \ncdr network model~\cite{dncr_model} to model the point-to-point communications and their predicted duration on the target topology with the given link characteristics.
Pfennig et al. verified this implementation in~\cite{haec_sim_verify}.
While HAEC-SIM uses sophisticated modeling of point-to-point communications (based on network coding), the model for collective communications adds a fixed minimum delay and serves as a temporal synchronization of all involved processes during simulation.

To predict the performance impact of the different mapping algorithms, we simulate the execution for each combination of application, mapping, and topology (see Table~\ref{tbl:parameters}).
We analyze the resulting HAEC-SIM application traces post-simulation. The configuration files used for the simulations can by found in Appendix~\ref{app:haec_config}

\todo[inline]{@Mario: Provide the configurations used. Mario: The config files should be an appendix to the arxiv version.}

\subsection{Post-simulation Performance Evaluation}\label{sub:postsim}

We use the simulation results (traces) to calculate the communication model time, the cost of MPI point-to-point communications, the overall parallel cost of the application~\cite{IntroPC:2015}, as well as post-simulation communication metrics: volume, distance, and dilation.

The communication model time is the \emph{sum} of the duration of all point-to-point message transfers on the transport layer according to the \ncdr communication model.
The MPI point-to-point cost denotes the \emph{aggregated} time that all application processes spent in MPI point-to-point communication functions, such as \texttt{MPI\_Send}, \texttt{MPI\_recv}, and \texttt{MPI\_Wait}.

\begin{figure*}[!htb]
	\centering
	\subfigure[CG]{
		\includegraphics[width=.46\linewidth]{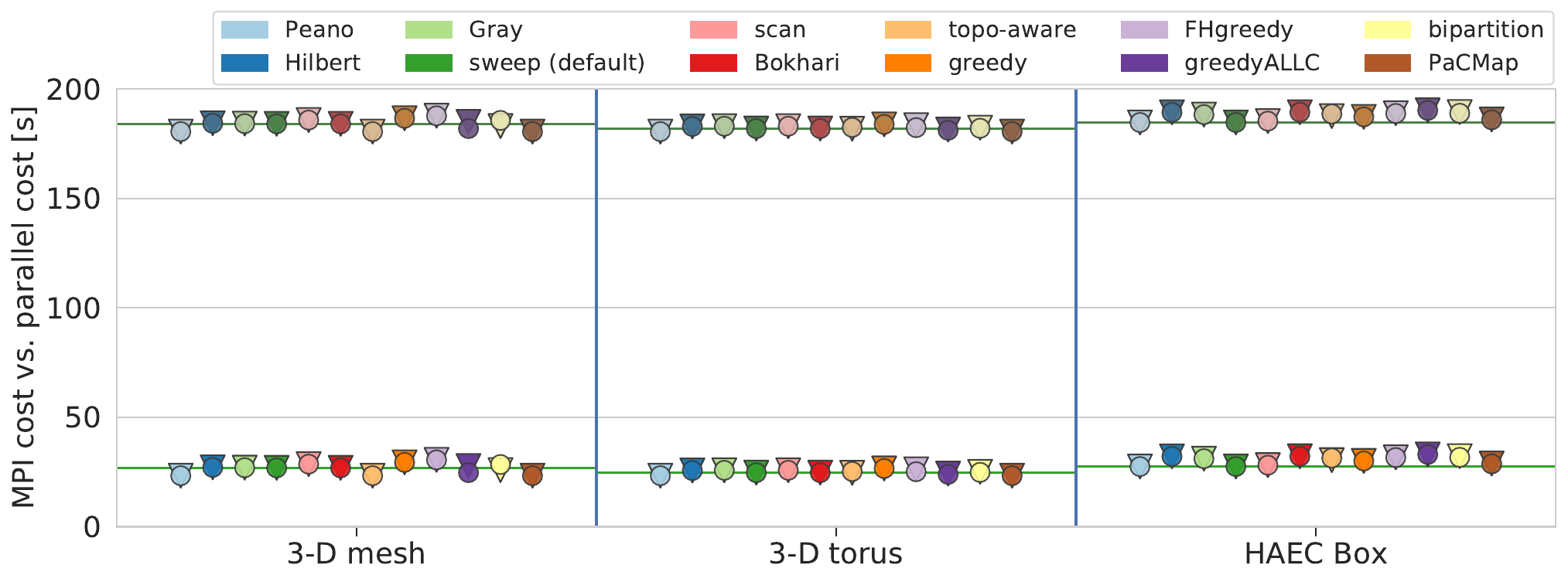}
		\label{subfig:cgaggp2p}
	}
	\subfigure[BT-MZ]{
		\includegraphics[width=.46\linewidth]{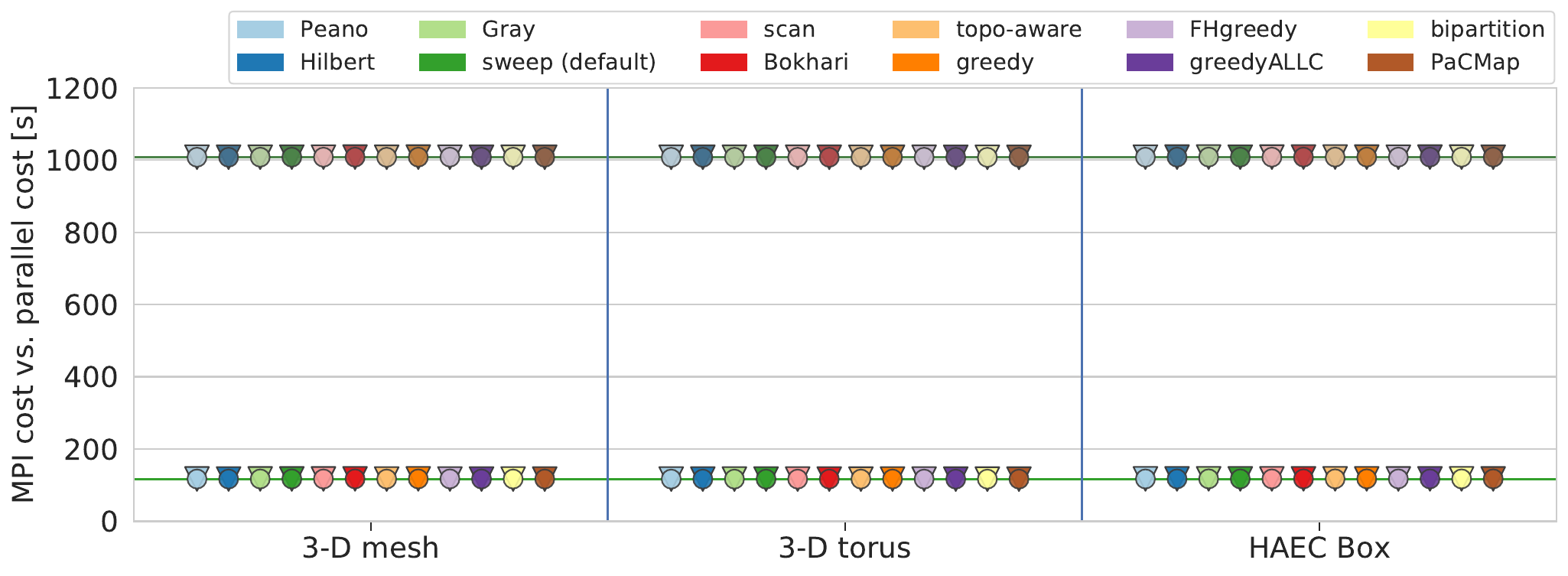}
		\label{subfig:btaggp2p}
	}
	\subfigure[LULESH]{
		\includegraphics[width=.46\linewidth]{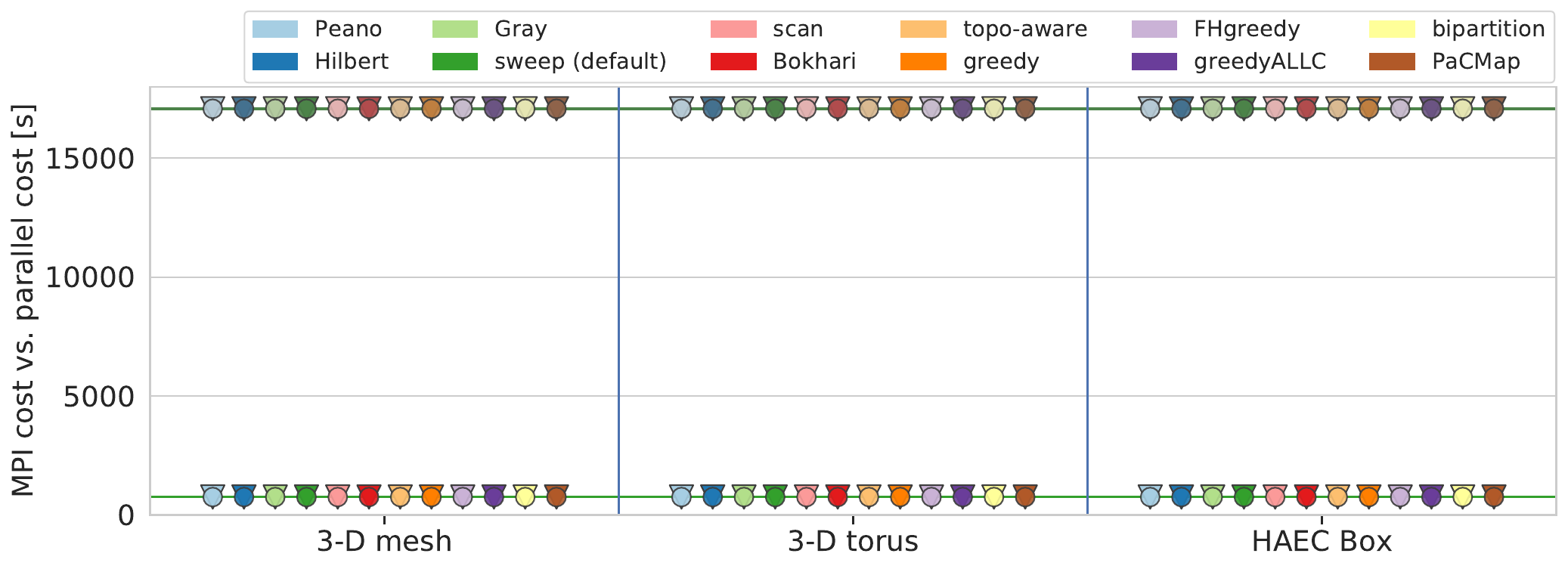}
		\label{subfig:luleshaggp2p}
	}
	\subfigure[AMG]{
		\includegraphics[width=.46\linewidth]{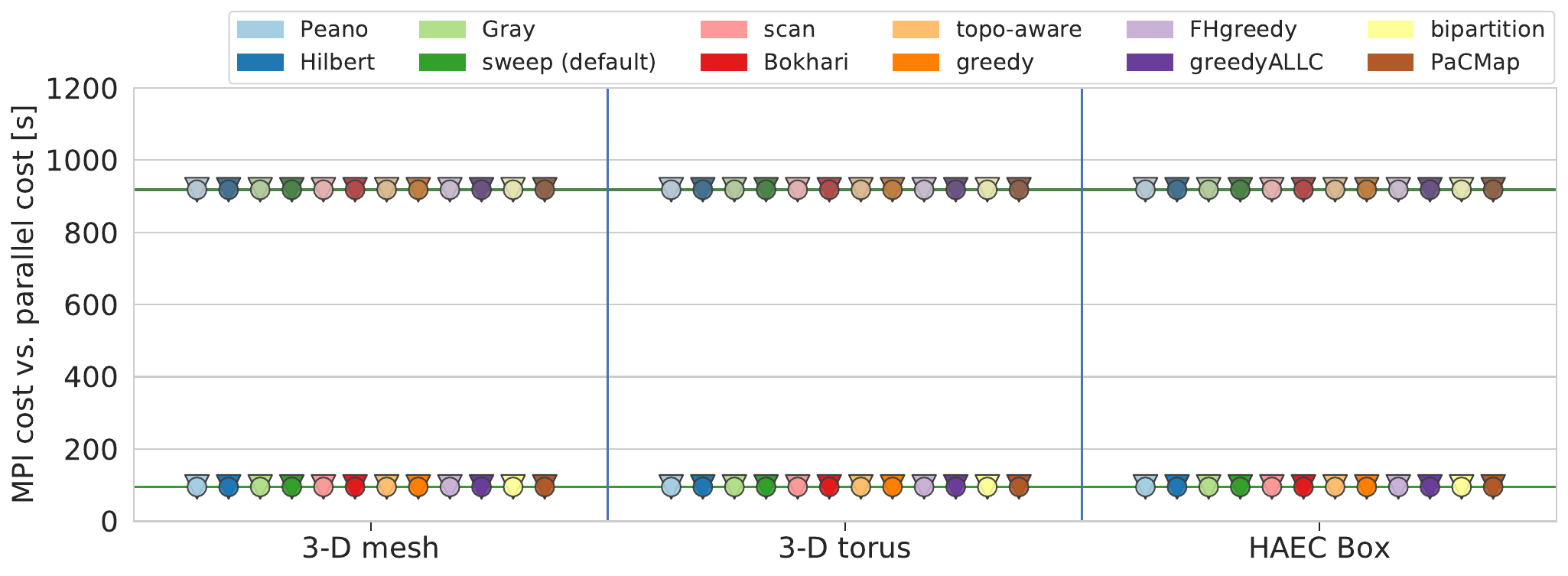}
		\label{subfig:amgaggp2p}
	}
	\caption[]{
		Simulated parallel execution cost and parallel communication cost for all applications, mappings, and topologies.
		Topologies are shown on the $X$ axis.
		The lower range on the $Y$ axis shows the MPI point-to-point cost while the upper range on the $Y$ axis shows the parallel cost.
		Each mapping is represented with a distinct color.
		Circles $\circ$ denote mappings based on \cmcount, while triangles $\triangledown$ indicate mappings based on \cmsize.
		The \textcolor{sweepgreen}{green} horizontal lines delineate the performance achieved by \textcolor{sweepgreen}{\texttt{sweep}} for both cost metrics.
	}
	\label{fig:aggtimes}
\end{figure*}

Fig.~\ref{fig:aggtimes} shows the impact of process mapping on application performance.
The plots present the values of the application's \emph{parallel cost} (calculated as parallel wall time multiplied by the number of nodes) and the part attributed to the \FC{\emph{aggregated}} MPI point-to-point costs.

%
\emph{These results show that, except for CG, neither the topology nor the mapping impact the values of the above performance metrics.}
For CG, the 3-D \texttt{torus} topology offers the highest performance, in particular with \texttt{Peano} and \texttt{PaCMap}.
Additionally, for the \mbox{3-D} \texttt{mesh} topology, the \texttt{topo-aware} mapping yields high performance.
On the \texttt{HAEC} \texttt{Box}, \texttt{sweep} has the lowest MPI and parallel costs.

\begin{figure*}[!htb]
	\centering
	\subfigure[CG]{
		\includegraphics[width=.46\linewidth]{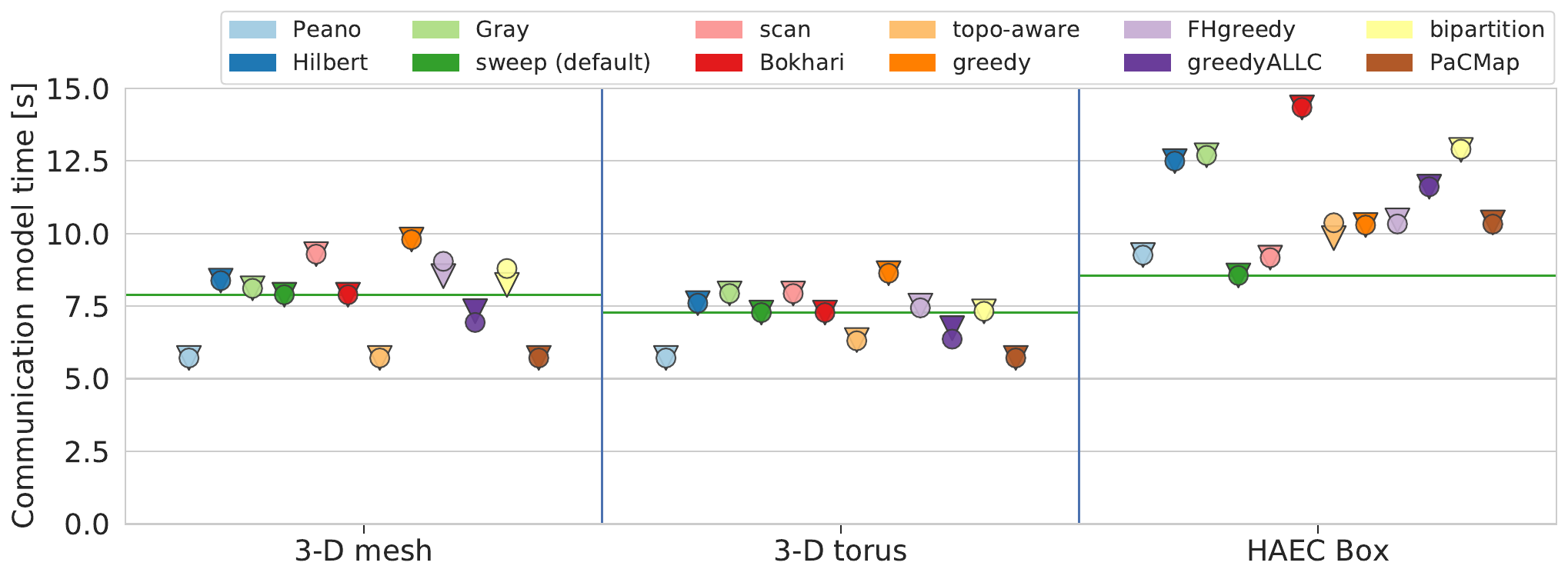}
		\label{subfig:cgagg}
	}
	\subfigure[BT-MZ]{
		\includegraphics[width=.46\linewidth]{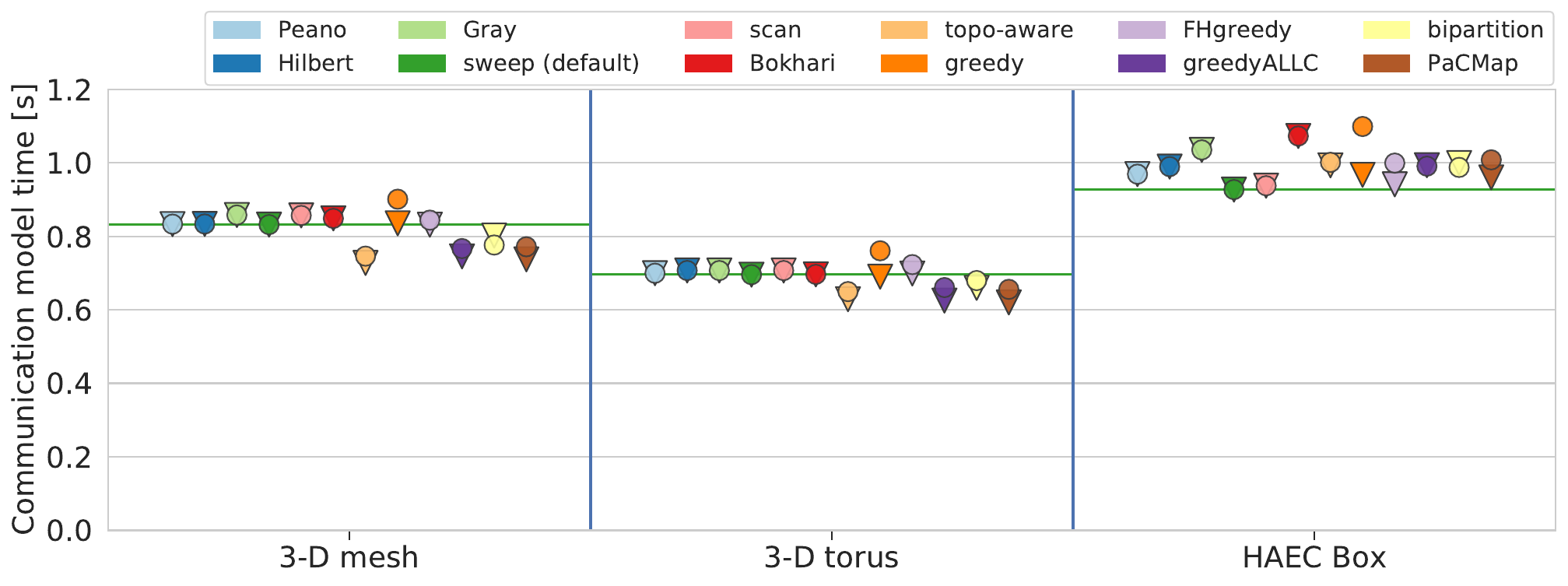}
		\label{subfig:btagg}
	}
	\subfigure[LULESH]{
		\includegraphics[width=.46\linewidth]{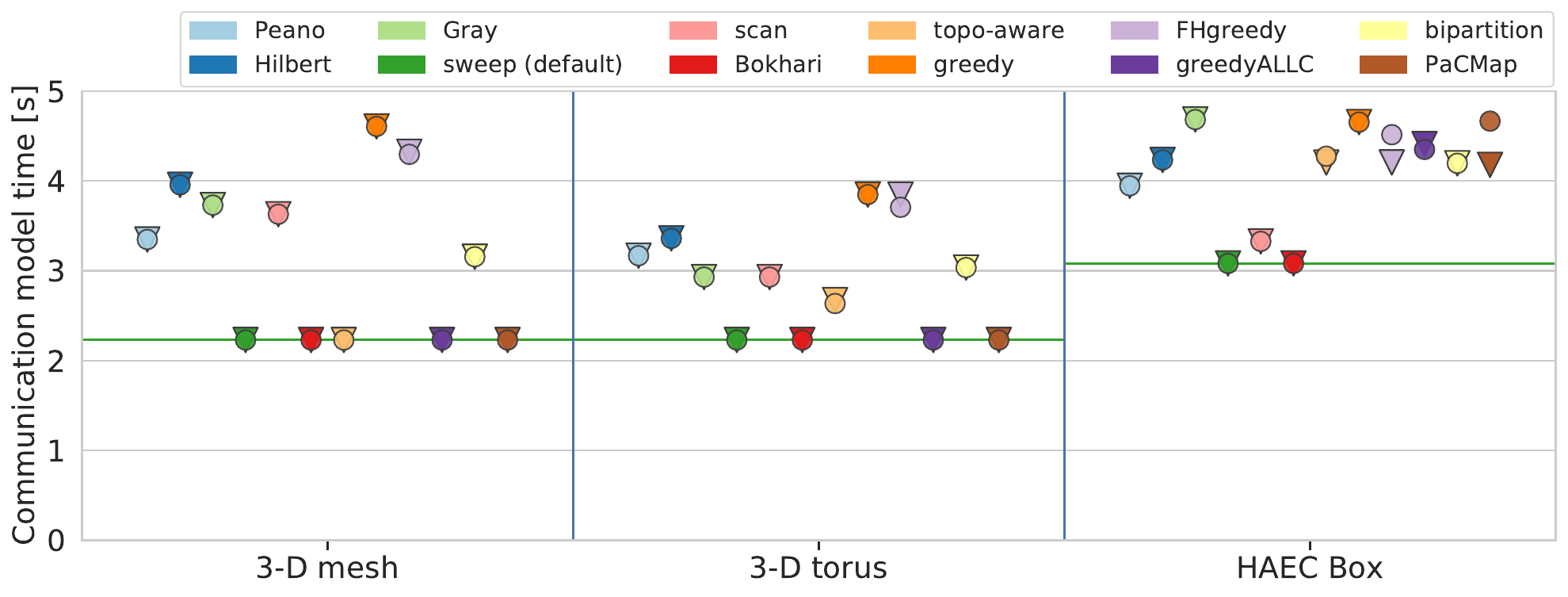}
		\label{subfig:luleshagg}
	}
	\subfigure[AMG]{
		\includegraphics[width=.46\linewidth]{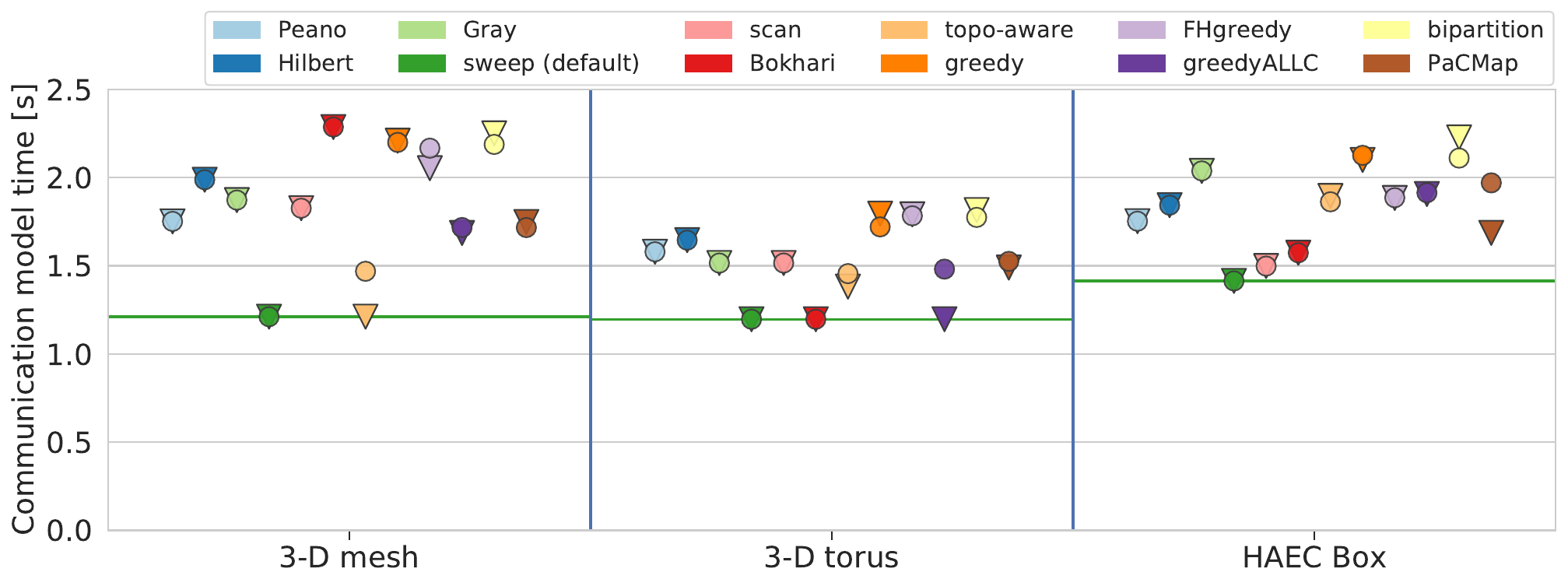}
		\label{subfig:amgagg}
	}
	\caption[]{
		Communication model time for all applications, mappings, and topologies.
		Topologies are shown on the $X$ axis.
		The $Y$ axis shows the sum of the transmission time over the transport layer according to the \ncdr model.
		The distinct colors represent the mappings.
		Circles $\circ$ denote mappings based on \cmcount, while triangles $\triangledown$ indicate mappings based on \cmsize.
		The \textcolor{sweepgreen}{green} horizontal lines delineate the performance achieved by \textcolor{sweepgreen}{\texttt{sweep}}.
	}
	\label{fig:commtime}
\end{figure*}

However, a different conclusion becomes visible, when we look at the communication model time in Fig.~\ref{fig:commtime}.
This figure shows the costs of the communication from the network perspective and reveals significant differences in the communication time.
One can notice that the \texttt{HAEC Box} shows the highest minimum times among the topologies in Fig.~\ref{fig:commtime}.
This is an effect of its wireless links that possess higher latencies (Table~\ref{tbl:links}).
In this situation, \texttt{sweep} and \texttt{scan} exhibit the highest performance as they map processes contiguously in the same $XY$ plane (Fig.~\ref{fig:sfcs}).
\emph{This observation highlights the need for process mapping algorithms tailored towards heterogeneous network topologies.}

While the mappings may not have influenced the application run-time in these simulations, \emph{the communication model time changed significantly.}
Therefore, at the network level, poor mappings lead to increased power consumption and a higher probability of contention, which are known to lead to significant performance loss.

\newpage

\subsection{Pre- and Post-simulation Performance Comparison}

Based on the knowledge about the application prior to simulation, we infer certain assertions about the predicted performance of the applications and use them to verify the simulation results.
According to the definition of dilation, its pre-simulation values cannot differ from its post-simulation values.
As part of the automated simulation process, we calculate this metric before and after each simulation experiment and assert that it remains unchanged.

All mapping algorithms use \cmcount and \cmsize as input (see Section~\ref{subsection:commMatrices}).
This duality results in two mappings for every application--mapping--system configuration.
Due to the nature of the communication-oblivious process mapping algorithms, we effectively generate the same mapping twice, i.e., the output based on \cmcount is equal to the one for \cmsize.
As the simulation is deterministic, the simulated performance is expected to be equal for these mapping tuples.
Fig.~\ref{fig:aggtimes} illustrates and confirms this, which increase the confidence in the results in both Fig.~\ref{fig:aggtimes} and Fig.~\ref{fig:commtime}.

In the \emph{pre-simulation} performance evaluation (Section~\ref{sub:presim}), we measure the \mbox{dilation} for the different applications--mappings--topologies configurations.
Given the significant differences in the dilation, this metric hints at great performance optimization opportunities.
However, except for the CG application, one can only see minor differences in the simulated parallel costs (Fig.~\ref{fig:aggtimes}).
This behavior stems from several intricacies, which are discussed in the following.

As shown in Table~\ref{table:compcomratio}, CG is the only application that frequently uses blocking MPI point-to-point operations.
In combination with the high-speed future-generation network links (see Table~\ref{tbl:links}), the other applications successfully mask their communication costs from being measurable from the application perspective by using non-blocking MPI point-to-point operations.
Moreover, the communication volume in CG is an order of magnitude higher than that of the other applications highlighting the effect of the communication masking even further.

Furthermore, since \ncdr is a \mbox{contention-oblivious}~communication model, concurrent communications do not content for resources and, therefore, no performance degradation can be observed without explicitly modeling congestion (part of future work).

The comparison of \emph{pre-simulation} dilation against the corresponding MPI point-to-point communication time, leads to new insights.
The communication times differences in Fig.~\ref{fig:commtime} are in agreement with the performance prediction derived from the dilation for all applications mapped to the 3-D \texttt{mesh} and \mbox{3-D} \texttt{torus} topology.
For the \texttt{HAEC Box} topology, no such correlation can be made.
\emph{This shows that the dilation metric needs a parameter to distinguish hops with different characteristics that are prevalent in heterogeneous networks.}

These observations confirm that \emph{dilation is a suitable metric for the prediction of communication time on homogeneous topologies}.
\emph{They also reveal the need for a novel performance metric for process mapping on heterogeneous topologies.}

%% file: 08-conclusion.tex
\section{Conclusions and Future Work}\label{sec:conclusion}

In this work we showed that process mapping is an important application optimization step that should be performed at every execution.
In a real setting, our work can be practiced first by binding the processing elements to their dedicated hardware processing units and subsequently by reordering the MPI ranks of the application according to the relevant mapping algorithm.

We proposed and exemplified a generic workflow to support process mapping as an explicit~application optimization step.
Using trace-based simulation, we predicted the communication performance for four applications, mapped with twelve mapping algorithms (implemented as a Python library), executing on three topologies.
Using pre- and post-simulation metrics, we showed that communication-aware mappings frequently outperform communication-oblivious ones.

We observed that dilation (measured as hop$\cdot$Byte) does not correlate well with the simulated application execution time.
However, when considering exclusively at the aggregated communication times, performance varies significantly among the mappings.
For heterogeneous topologies, the dilation metric would need a parameter to distinguish hops with different characteristics.
This observation indicates that even if the application execution time does not change, a careful mapping reduces the network load.
Therefore, \emph{mapping matters}.

In future work, 
the \maplib library can be augmented with other mapping algorithms, such as those from Bhatele et al.~\cite{hopbyte2010automate} and Wu et al.~\cite{TopoMap:2017}.
We see that HAEC-SIM could also be integrated with TopGen~\cite{TopGen:2018} to simulate additional interconnection networks and a contention-aware communication model.
Experiments on real systems are also part of future work, as we currently do not have access to the studied topologies at the moment.
Inter-application interference and the impact of process mapping on other metrics (e.g., congestion) are also part of future work.
A scalability study considering topologies with higher dimensions and a greater number of processes is also an important aspect to be considered going forward.
Finally, the impact of the combination of application-aware routing algorithms with communication-aware mapping algorithms is an interesting research question to pursue in the future.

%% file: 09-acknowledgment.tex
\section*{Acknowledgments}
This work is in part supported by the Swiss National Science Foundation in the context of the ``Multi-level Scheduling in Large Scale High Performance Computers'' (MLS) grant, number 169123 and by the German Research Foundation (DFG) within the CRC912 - HAEC.
The authors acknowledge Daniel Besmer and Viacheslav Sharunov for their earlier contribution to this work.


%% file: 10-appendix.tex
\begin{appendix}

\section{Appendix}

\subsection{HAEC-SIM configuration files}\label{app:haec_config}

\begin{lstlisting}[language=json,caption={HAEC-SIM configuration file for the 3-D \texttt{mesh} topology},label=lst:config_mesh]
{
  "platform": {
    "link_types": [
      {
        "name" : "optical",
        "bandwidth" : 31250000000,
        "latency" : 10,
        "bit_error_rate": 1e-12,
        "packet_attack_rate": 0.01
      },
      {
        "name" : "wireless",
        "bandwidth" : 12500000000,
        "latency" : 100,
        "bit_error_rate": 1.0e-8,
        "packet_attack_rate": 0.01
      }
    ],
    "topology": {
      "shape": [ 4, 4, 4 ],
      "link_types": [ "optical", "optical", "optical" ],
      "connectivity": [ "mesh", "mesh", "mesh" ]
    }
  },
  "modules": {
    "static_network_model": {
      "routing": "shortest_path",
      "parameter_search_folders": "haec-sim/share/modules/static_network_model/parameters",
      "communication_model": "DNCr",
      "size_packet": 1500,
      "size_finitefield": 8,
      "size_window": 5,
      "delay_processing": 63,
      "delay_mpi": 5,
      "size_mpi_header": 16,
      "size_windowid": 4,
      "size_packetid": 2,
      "size_signature": 256,
      "size_generationid": 4
    }
  }
}
\end{lstlisting}
\begin{lstlisting}[language=json,caption={HAEC-SIM configuration file for the 3-D \texttt{torus} topology},label=lst:config_torus]
{
  "platform": {
    "link_types": [
      {
        "name" : "optical",
        "bandwidth" : 31250000000,
        "latency" : 10,
        "bit_error_rate": 1e-12,
        "packet_attack_rate": 0.01
      },
      {
        "name" : "wireless",
        "bandwidth" : 12500000000,
        "latency" : 100,
        "bit_error_rate": 1.0e-8,
        "packet_attack_rate": 0.01
      }
    ],
    "topology": {
      "shape": [ 4, 4, 4 ],
      "link_types": [ "optical", "optical", "optical" ],
      "connectivity": [ "torus", "torus", "torus" ]
    }
  },
  "modules": {
    "static_network_model": {
      "routing": "shortest_path",
      "parameter_search_folders": "haec-sim/share/modules/static_network_model/parameters",
      "communication_model": "DNCr",
      "size_packet": 1500,
      "size_finitefield": 8,
      "size_window": 5,
      "delay_processing": 63,
      "delay_mpi": 5,
      "size_mpi_header": 16,
      "size_windowid": 4,
      "size_packetid": 2,
      "size_signature": 256,
      "size_generationid": 4
    }
  }
}
\end{lstlisting}
\begin{lstlisting}[language=json,caption={HAEC-SIM configuration file for the 3-D \texttt{haec} topology},label=lst:config_haec]
{
  "platform": {
    "link_types": [
      {
        "name" : "optical",
        "bandwidth" : 31250000000,
        "latency" : 10,
        "bit_error_rate": 1e-12,
        "packet_attack_rate": 0.01
      },
      {
        "name" : "wireless",
        "bandwidth" : 12500000000,
        "latency" : 100,
        "bit_error_rate": 1.0e-8,
        "packet_attack_rate": 0.01
      }
    ],
    "topology": {
      "shape": [ 4, 4, 4 ],
      "link_types": [ "optical", "optical", "wireless" ],
      "connectivity": [ "torus", "torus", "crossbar" ]
    }
  },
  "modules": {
    "static_network_model": {
      "routing": "haec_box",
      "parameter_search_folders": "haec-sim/share/modules/static_network_model/parameters",
      "communication_model": "DNCr",
      "size_packet": 1500,
      "size_finitefield": 8,
      "size_window": 5,
      "delay_processing": 63,
      "delay_mpi": 5,
      "size_mpi_header": 16,
      "size_windowid": 4,
      "size_packetid": 2,
      "size_signature": 256,
      "size_generationid": 4
    }
  }
}
\end{lstlisting}

\end{appendix}